\documentclass[superscriptaddress,secnumarabic,amssymb,nobibnotes,prfluids]{revtex4-2}

\usepackage{subfig}
\usepackage{float}
\usepackage{setspace}
\usepackage{graphicx}
\usepackage{natbib}
\usepackage{rotating}
\usepackage{amsmath}

\usepackage{tikz,pgfplots}
\usetikzlibrary{patterns}
\usetikzlibrary{calc}
\usetikzlibrary{arrows,shapes}
\usetikzlibrary{positioning}

\newlength{\FigureHeight}
\newlength{\FigureHeightHalf}
\newcommand{\FigureXYLabel}[5]{%
\settoheight{\FigureHeight}{#1}%
\setlength{\FigureHeightHalf}{0.5\FigureHeight}%
\begin{center}%
\raisebox{\FigureHeightHalf}{\makebox{#4\makebox[#5]{}}}%
#1\\%
\vspace{#3}%
#2\\%
\end{center}%
}

\begin{document}

\title{Aerodynamic sound of turbulent time-developing shear layer as
the outcome of the flow linear non-modal instability}

\author{George Khujadze\footnote{Corresponding author: george.khujadze@uni-siegen.de}}%
\affiliation{Chair of Fluid Mechanics, Universit\"at Siegen, 57068 Siegen, Germany}

\author{David Gogichaishvili}
\affiliation{Department of Physics, The University of Texas at Austin, Austin 78712, USA}

\author{George Chagelishvili}
\affiliation{E. Kharadze Georgian National Astrophysical Observatory, Abastumani 0301, Georgia}%
\affiliation{Institute of Geophysics, Tbilisi State University, Tbilisi 0128, Georgia}%

\author{Alexander Tevzadze}%
\affiliation{E. Kharadze Georgian National Astrophysical Observatory, Abastumani 0301, Georgia}%
\affiliation{Kutaisi International University, Kutaisi 4600, Georgia}%

\author{Jan-Niklas Hau}%
\affiliation{Technische Universit\"at Darmstadt,   Darmstadt, 64287, Germany%
\footnote{The main contribution of this author happened when still at Darmstadt Technical university. {\it Present address:} Frankfurt, Germany}}

\author{Holger Foysi}%
\affiliation{Chair of Fluid Mechanics, Universit\"at Siegen, 57068 Siegen, Germany}

\date{\today}

\begin{abstract}
The aim of this paper is to substantiate the importance of
non-normality of shear flow linear operators and its consequence --
the non-modal dynamics of the perturbations -- in the formation of
acoustic wave output of time-developing free shear/mixing layers.
Initially, the linear transient dynamics of spatial Fourier harmonics is
considered in a 3D homentropic parallel unbounded inviscid constant
shear flow which can model the central/body part of the shear layer.
The non-modal approach allows to capture the only linear mechanism
of the acoustic wave generation -- \textit{the linear vortex--wave
mode coupling induced by the shear flow non-normality}. We describe
the specific/key features of the generation process that should
leave traces on the further dynamics of the generated waves.
Thereafter, the results of direct numerical simulations of
compressible turbulent time-developing mixing layers for a moderate
convective Mach number (specifically, $M_c=0.7$) and simulation
boxes $(L_x,L_y,L_z)$ with fixed streamwise and shearwise lengths
($L_x=100, L_y=200$) and different streamwise-spanwise aspect ratios
($L_x/L_z=0.5,1,2$) are presented. The simulations identify the
origin of the acoustic wave output: the dominance of a \emph{linear
generation process of acoustic waves in the shear layer core region, induced
by the flow non-normality}, observable in the near field of
acoustic waves emitted by the flow.

The non-modal physics of the perturbation dynamics studied here, utilizing
the example of a time developing compressible turbulent mixing
layer, must be essential also for other turbulent engineering shear
flows as the non-normality is their inherent feature. This will be relevant
also for engineering compressible shear flows with exponentially
growing ``modal solutions'', whose finite time period
dynamics/growth undergoes a significant change due to the non-modal
physics. The consideration of the non-modal effects (that has been
successfully adopted in hydrodynamic, atmospheric and astrophysical
flow communities since the 1990s) is necessary in order to properly describe
the flow's real dynamics and, most importantly in the
context of our study, to get physical insight into the formation of
the near acoustic field.
\end{abstract}

\maketitle

\section{Introduction}
Aerodynamic sound generation is a major subject of fluid dynamics
research and represents an enormous scientific and technological
challenge. A simplified model to get physical insight into the
mechanisms of aerodynamic sound generation in jets and other
turbulent flows is the compressible free shear or mixing layer -- the flow
consisting of two parallel streams of fluid with unequal velocities -- as it
mimics shear layer regions of some natural and engineering
nonuniform flows. For quite some time already it became clear that
the flow turbulence is a mix of chaotic and coherent motions,
focusing interest on coherent structures in search of aerodynamic
sound sources. According to the formed view, the main sources of the
aerodynamic sound emitting from shear layers are large-scale
structures in the form of wave-packets -- perturbative structures
that are spatio-temporally coherent over distances and times usually larger
than the integral scales of the turbulence. The dominance of these
wave-packet-type sources of sound for supersonic shear layers was
shown in \cite{Tam1995}. In the case of subsonic shear layers, there
are strong experimental, theoretical and computational evidences of
the wave-packets effectiveness as the source of most of the sound
output (see, e.g., \cite{Laufer1983,Crighton1990,Cheung2009,Colonius1997}).
The importance of wave-packets in acoustics can be learned from
\cite{Jordan2013}, where their existence, energetics, dynamics and
acoustic efficiency for jet flow configurations is exhaustively
reviewed. On the basis of the extensive number of simulations and
modern measurement techniques discussed in there, wave-packets are represented
``as instability waves, or more general
\emph{modal solutions} of the governing equations'', confirming
their acoustic importance in high subsonic and supersonic jets.
Highlighting the phrase ``modal solution'' by us is intentional, as
this concept is the turning point of our analysis. Specifically,
it turned out that the traditional modal analysis of the
perturbation dynamics in shear flows is not optimal and can be
misleading in many respects.

Shortcomings of the traditional modal analysis of linear processes
in shear flows (the spectral expansion of perturbations in time with
subsequent analysis of its eigenmodes) were rigorously revealed by
the hydrodynamic community in the 1990s (see, e.g.,
\cite{Gustavsson1991,Butler1992,Trefethen1993,Reddy1993a,Reddy1993b,Trefethen1993,Schmid2001,Schmid2007}).

Operators with nonorthogonal eigenfunctions are referred to as
non-normal \cite{Trefethen1997,Trefethen2005a}. Consequently, flows
with such operators are referred to as non-normal. It has been shown
that the operators of the modal analysis of linear processes in
shear flows are exponentially far from normal
\cite{Schmid2001,Schmid2007,Henningson1994} (the norm of the flow
non-normality increases with the shear rate). Due to this
non-normality, the eigenmodes strongly interfere and a suitable
approach needs to be applied in order to fully analyze this behavior
- the analysis of eigenmodes, treated as linearly independent, is
misleading. In the framework of the classical modal approach, that
focuses on the asymptotic stability of flows, little attention is
paid to any particular initial value or finite time period of the
dynamics. Consequently, the transient behavior, that is the
characteristic feature of the dynamics of shear flows, is considered as
non-significant. By discarding the modal analysis and adopting the
so-called non-modal approach to aerodynamic sound generation,
manifestations of the flow non-normality can be traced in the
acoustic wave output of, for instances, time-developing mixing layers
(with streamwise and spanwise periodic boundary conditions). The non-modal
approach reveals a remarkably rich picture of the linear
perturbation behavior -- e.g., vortex--wave mode coupling -- that
greatly differs from solutions of the modal approach.
Mathematically, the linear vortex--wave mode coupling has been
rigorously studied. Taking the specifics of the linear coupling into
account, numerous examples of spontaneous wave generation by vortex
mode disturbances in atmospheric, astrophysical and magnetized
smooth shear flows are already studied and described in detail
\cite{Tevzadze1998,Tevzadze2003,Heifetz2003,Vanneste2004,Kalashnik2006,
Vanneste2008,Olafsdottir2008,Lott2010,Mamatsashvili2010,Favraud2013,Vanneste2013}.
As for the acoustic wave generation, the coupling has been analyzed
numerically for mode-dependent moderate and large Mach numbers
\cite{Chagelishvili1997,Farrell2000,Favraud2014,Hau2015}, as well as
in terms of Stokes lines for small Mach numbers \cite{Bakas2009}.

Here, the following should also be emphasized. Generally, the shear
flow non-normality essentially changes the finite time period of the
dynamics. This concerns not only the appearance of the new type of
linear transient phenomenon, the vortex--wave mode coupling, but
also relates to exponentially growing ``modal solutions'', whose
initial, finite time period dynamics/growth undergoes a significant
change. It is necessary to take into account the non-modal effects
on the exponentially growing mode dynamics instead of relying solely on its
modal growth, in order to properly describe their real dynamics.
This view about the importance of non-modal physics in the
dynamics of exponentially growing modes is well comprehended by the
astrophysical disks community
\cite{Squire_Bhattacharjee14,Gogichaishvili18}. Astrophysical disks
are highly non-normal sheared flows due to their strong differential
rotation. Consequently, the flow non-normality is investigated in
all its manifestations and one can use the experience and knowledge
of the astrophysical disk community for the investigation of the
sound production in engineering turbulent shear flows. One example of
the necessity of making use of this description can be given. The modal approach
describes exponentially/spectrally unstable modes of wave-packets in
jet flows \citep{Jordan2013}. This description can be misleading, however,
without involving the flow non-normality modifications of the
spectrally unstable mode dynamics.

In this paper we consider the compressible mixing layer, a simplified,
and thus, convenient model to provide physical insight into the
mechanisms of aerodynamic sound generation in jet and other
turbulent engineering flows. This flow mimics shear layer regions of
jets and, furthermore, by itself, occurs in many natural and
engineering flows, being therefore just sophisticated enough to
observe the relevant physics of interest, here. The object of our
study is the time-developing mixing layer, because the numerically
simulated fully nonlinear/turbulent dynamics of this flow enables us
to follow quite clearly the outcomes of the vortex--wave linear
coupling in smooth shear flows. Indeed, the central/core part of
the time-developing mixing layer (i.e. the area between the near
boundary/turning areas) can be modeled as a constant shear flow.
This central part of the flow is energetically active due to the
non-normality: this ensures linear transient growth of certain
Fourier harmonics of vortex mode perturbations and linear generation
of acoustic wave harmonics by related vortex mode ones. The boundary
areas  modify the dynamics of the central part in a certain way.
However, the latter still enables to grasp the phenomena that appear
in linearly sheared regions and can therefore form the basis of
far-field sound of more complex nonuniform flows.

First of all, we investigate the linear generation of acoustic waves
by vortex modes in a 3D homentropic parallel unbounded inviscid
constant shear flow: ${\bf U}_0=(Ay,0,0)$, where the flow is
directed along the $x$-axis, the flow shear with parameter $A$ -- along
$y$ axis. We consider the dynamics of a single  pure vortex mode
Fourier harmonic ($k_x,k_y(t),k_z$).  According to the non-modal
approach the shearwise wavenumber of harmonics  depends linearly on
time, $k_y(t)=k_y(0)-Ak_xt$. The linear generation of acoustic waves
and its characteristics are grasped by studying initially highly
up-shear tilted pure vortex mode Fourier harmonics, $k_y(0)/k_x \gg 1$,
independent of the value of the spanwise wavenumber, $k_z$.
The generation process of the acoustic wave harmonics is specific and occurs
due to linear transient vortex--wave mode coupling, induced by the shear flow
non-normality. While the potential vorticity of the
imposed vortex mode harmonic is preserved in time, a wave part
appears just when the time-dependent shearwise wavenumber of the
analyzed harmonic becomes zero. The key to comprehending the
mathematical aspect of the wave generation is rooted in the possibility of
splitting the velocity field of the considered harmonic into its
vortex and wave parts at exactly this time. The potential vorticity
of the harmonic is fully contained in the vortex part, which evolves
aperiodically when tilted down-shear $(k_y/k_x<0)$. At the same time, the
down-shear tilted wave part has a zero potential vorticity and
exhibits an oscillating nature in time. The performed analysis
reveals the key characteristics of acoustic wave harmonics' generation
process: increase of the efficiency of the wave generation with the
increase of streamwise length scale of the vortex harmonics; abrupt
emergence of the wave harmonics from the vortex ones; regular/fixed
phase of the wave harmonics at the moment of emergence (that is the
basis of constructive interference of the linearly generated
harmonics). These key characteristics are due to the specific nature
of the linear vortex--wave mode coupling induced by shear flow non-normality.

Next we perform direct numerical simulations (DNS) of compressible
turbulent time-developing mixing layers for moderate convective Mach
number, $M_c= 0.7$, and simulation boxes, $(L_x,L_y,L_z)$, for fixed
streamwise and shearwise lengths ($L_x=100, L_y=200$) and different
streamwise-spanwise aspect ratios ($L_x/L_z=0.5,1,2$). The size of
the computational boxes in the shear direction, $L_y$, exceeds the
shear layer thickness a few times, allowing to grasp the acoustic
wave output (\emph{near wave field}) of the layer and its
characteristics.
The choice of the convective Mach number and the streamwise and
spanwise lengths of the boxes mainly make the harmonics with the
largest streamwise wavelength ``acoustically active'', ensuring the
identification of the origin of the near wave field, emitted
mostly because of the only linear mechanism of the acoustic wave
generation in the central part of the flow induced by its
non-normality \footnote{To note is that we focused on the acoustic
wave linear generation due to the linear vortex--wave mode coupling
induced by shear flow non-normality. However, generally there exists
another linear mechanism of the wave generation, which is also
induced by the flow non-normality but not through the vortex mode --
this mechanism leads to the direct extraction of the flow energy by
acoustic waves \cite{Chagelishvili1994,Chagelishvili1997c}. The
growth of the acoustic wave harmonics is algebraic (i.e., not
exponential), therefore it needs time to achieve palpable values to
make any contribution to the near acoustic wave field. This is
not a case for the considered time-developing shear layer --
acoustic wave harmonics don't have enough time to extract mean shear
flow energy directly and amplify. Therefore, we can not observe this
process in our simulations. In this connection, the acoustic wave
linear generation due to the vortex--wave mode coupling in the paper
is referred to as the only mechanism.}.
The dominance of this linear mechanism in the aerodynamic sound
generation points out the shortcomings of the traditional modal
analysis of linear processes in shear flows. The proposed view is in
accordance with the established picture about the decisive role of
large-scale structures in shear layers (i.e. wave-packets) in
aerodynamic sound radiation. However, it diverges with it by the
features of the linear generation of acoustic waves: shear flows are
nonnormal and the generation analysis in the framework of the modal
approach is misleading unless analyzed in the framework of the
non-modal approach. The basic/simplest ``elements'' of aerodynamic
sound generation are considered not entire and undivided wave
packets, but their ``building blocks'': vortex mode Fourier
harmonics, or, differently, Kelvin modes.

The Numerical simulations in this paper show a mostly 2D regular
character of the acoustic wave field output at $L_x/L_z=2$; the
appearance of slight 3D features of the output at $L_x/L_z=1$ and a
further significant increase of the output irregularity at
$L_x/L_z=0.5$, that forms complex (almost chaotic) picture
containing a significant 3D component of the near acoustic
field. In reality, all these are in exact accordance with the above
mentioned linear transient vortex--wave mode coupling induced by the
shear flow non-normality.

The structure of the paper is as follows:
In section \ref{sec:model_equation} we outline the linear generation
of acoustic waves by vortex modes in the constant shear flow
focusing on key features (abrupt and regular character) of the
emergence of acoustic wave harmonics from the vortex ones and on the
dependence of the wave generation efficiency on the ratio $k_z/k_x$.
In section \ref{sec:DNS} the results of DNS of turbulent
time-developing shear layer are presented and the acoustic wave
field output of the shear layer is analyzed to show the decisive
role of the linear aerodynamic sound
generation induced by the flow non-normality in the forming of the far-field.
Summary and discussions are given in section \ref{sec:conclusion3D}.

\section{The linear generation of acoustic waves by vortex modes
in constant shear flows}
\label{sec:model_equation}
The central region of time-developing mixing layer can be modeled by
constant shear flow that is energetically active due to the
non-normality: it ensures a linear transient growth of certain
harmonics of the vortex mode perturbations and the linear generation
of acoustic wave harmonics by related vortex modes
\cite{Chagelishvili1997,Farrell2000,Favraud2014,Hau2015}. Of course,
dynamical processes in the central region of the shear layer are
influenced, in reality, by the processes in the boundary areas of
the flow. However, the linear transient processes can give birth to
acoustic waves in the central region (approximated by constant shear),
which form the significant (even dominant) portion of the near wave field of the flow.

Here, we outline the essence of the acoustic wave linear generation
process in 2D and 3D parallel inviscid constant shear flow,
with uniform density and pressure, $U_0(Ay,0,0)$, $\rho_0$,
$P_0=\rho_0c_s^2$, respectively, with $c_s$ being the speed of sound.
Employing the standard method of the non-modal approach
the spatial Fourier harmonics  of perturbations with a
time-dependent shear-wise wavenumber, $k_y(t)$,
are introduced \cite{Chagelishvili1997,Hau2015}:
$$
\Psi(x,y,z,t)= \Psi(k_x,k_y(t),k_z,t)e^{i k_xx+k_y(t)y+k_zz},
$$
\noindent where $k_y(t)=k_y(0)-k_xAt$. $\Psi \equiv
(u_x,u_y,u_z,\rho,p)$ denotes the velocity components, density and
pressure of spatial Fourier harmonics, respectively. In physical
space $(x,y,z)$, the above representation describes a harmonic plane
wave -- so-called, Kelvin mode -- with time-dependent amplitude and
shearwise wavenumber. The variation of $k_y(t)$ is caused by the
shearing of the harmonics due to the background flow.

The physical variables are normalized as follows
\begin{align}\label{nondim}
&{\bf v} = \frac{\mathbf {u}}{c_s}, ~~ D=i\frac{\rho}
{\rho_0}=i\frac{p} {P_0},
~~ \tau=c_s k_xt, ~~ {\mathcal {M}}=\frac{A}{k_x c_s }, \nonumber ~\\
&\gamma = \frac{k_z}{k_x}, ~~~\beta(\tau)= \frac{k_y(0)}{k_x} -
{\mathcal {M}}\tau= \beta(0) - {\mathcal {M}}\tau,
\end{align}
where ${\mathcal {M}}$ is the normalized shear rate and
characterizes compressibility of the 2D harmonics with streamwise
wavenumber $k_x$. We call ${\mathcal {M}}$ the mode-dependent
Mach number of a 2D harmonic.

3D harmonics have a different measure of compressibility. Therefore,
we separately introduce the mode-dependent Mach number at an
arbitrary moment of time, $\mathcal{M}_D(\tau)$, as the ratio of the
normalized shear rate, ${\mathcal {M}}$, to the normalized instant
frequency of the harmonic,
${\omega(\tau)}={\sqrt{1+\beta^2(\tau)+\gamma^2}}:$
\begin{equation}\label{eq:M_D}
\mathcal{M}_D(\tau) \equiv  \frac{\mathcal{M}}{\omega(\tau)} =
\frac{\mathcal{M}}{\sqrt{1+\beta^2(\tau)+\gamma^2}}
\end{equation}
The maximum of $\mathcal{M}_D$ is reached at the critical time
$\tau^\ast = \beta(0)/\mathcal{M}$, i.e., at the moment the Fourier
harmonics cross the $k_x$-axis (the line $k_y=0$). At this
point the measure of compressibility of a 3D mode reaches its maximum,
defined as:
\begin{equation}
\mathcal{M}^\ast \equiv  \max(\mathcal{M}_D(\tau) ) =
\frac{\mathcal{M}}{\sqrt{1+\gamma^2}}
\label{eq:M_max}
\end{equation}
This mode-dependent Mach number determines the linear mode-coupling
for the related 3D harmonic and, consequently, the efficiency of the
wave generation. The above equation shows that $\mathcal{M}^\ast$
has a maximum for the 2D harmonics (when $\gamma=0$) and decreases with
increase of $\gamma$, consequently, the wave generation should be
maximal for 2D harmonics and should decrease with an increase of $\gamma$.
These facts are confirmed by the calculations below.

At $\beta(0) \gg 1$, when $\mathcal{M}_D \ll 1$, the compressibility
characteristics of the harmonic substantially diminishes and the
separation between acoustic waves and vortex motion becomes
possible. Hence, we determine vortex mode harmonics in these regions
similar to \citep{Hau2015} in order to study the generation of
oscillating and propagating wave mode perturbations by initially
pure (void of wave modes) vortex modes. By extracting the vortex
part in regions of $\beta(0) \gg 1$ and subsequently exploiting the
properties of vortex modes, one can find harmonics of pure vortex
perturbations in any region/point of the wavenumber space.

The considered constant shear flow system contains an essential,
temporally conserved parameter for each harmonic called the
potential vorticity
\begin{equation}
\mathcal{W} = \left(1+\gamma^2\right)v_y - \beta(\tau)(v_x+\gamma v_z) - \mathcal{M} D.
\end{equation}
The vortex mode harmonics has non-zero potential vorticity and zero
group velocity, whereas the acoustic wave mode harmonics give rise
to zero potential vorticity and non-zero group velocity. By
initially imposing highly up-shear tilted pure vortex mode harmonics
($\beta(0) \gg 1$) in the linearized equations, the emergence of the acoustic wave
harmonics by the linear vortex--wave mode coupling caused by the
shear flow non-normality was investigated in \cite{Chagelishvili1997,Hau2015}.
The efficiency of this mechanism depends on \emph{the harmonic's
(mode-dependent) Mach number}. The wave generation for 2D harmonics
is noticeable at ${\mathcal {M}}=0.2$ and substantial already at
${\mathcal {M}}>0.3$. We would like to emphasize that we operate in
terms of two different definitions of Mach numbers:
a \emph{mode-dependent Mach number} (${\mathcal {M}}$ for 2D harmonics
and $\mathcal{M}^\ast$ for the 3D case) and a \emph{convective Mach number}
($M_c$). The first definition is related to separate Fourier harmonics
of perturbations and the second determines the flow compressibility
effects in general (in our case that of a time-developing mixing layer).

\begin{figure}
\begin{minipage}[c]{.26\linewidth}
\FigureXYLabel{\includegraphics[width = 1.\linewidth]{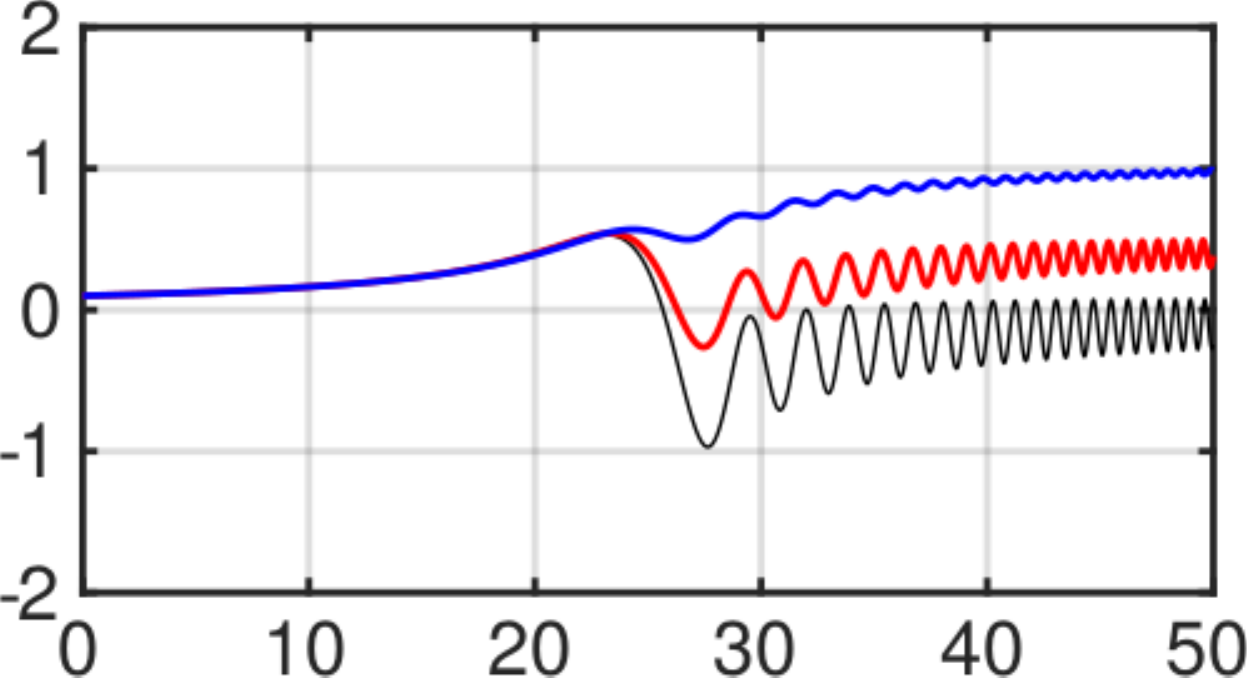}}
{$ $}{-3mm}{\begin{rotate}{90} {\small $v_{x}$} \end{rotate}}{1mm}
\end{minipage}
\hspace{0.5cm}
\begin{minipage}[c]{.26\linewidth}
\FigureXYLabel{\includegraphics[width = 1.\linewidth]{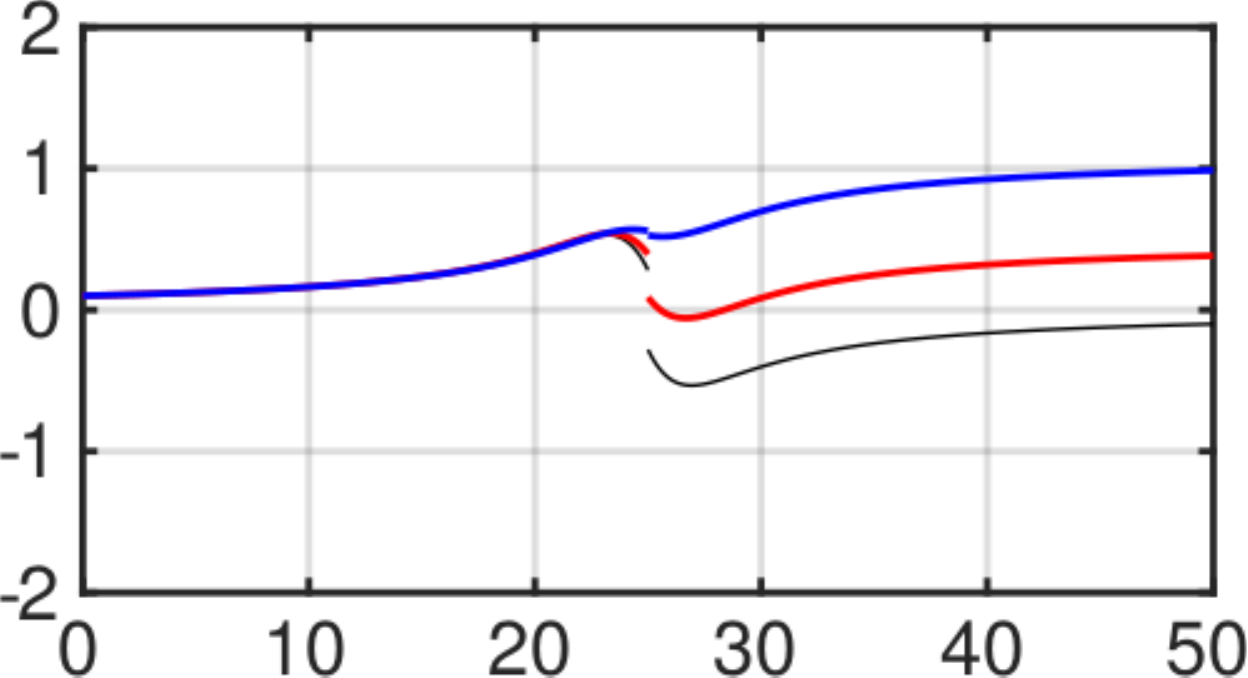}}
{$ $}{-3mm}{\begin{rotate}{90}$v_{x}^{(v)}$\end{rotate}}{1mm}
\end{minipage}
\hspace{0.5cm}
\begin{minipage}[c]{.26\linewidth}
\FigureXYLabel{\includegraphics[width = 1.\linewidth]{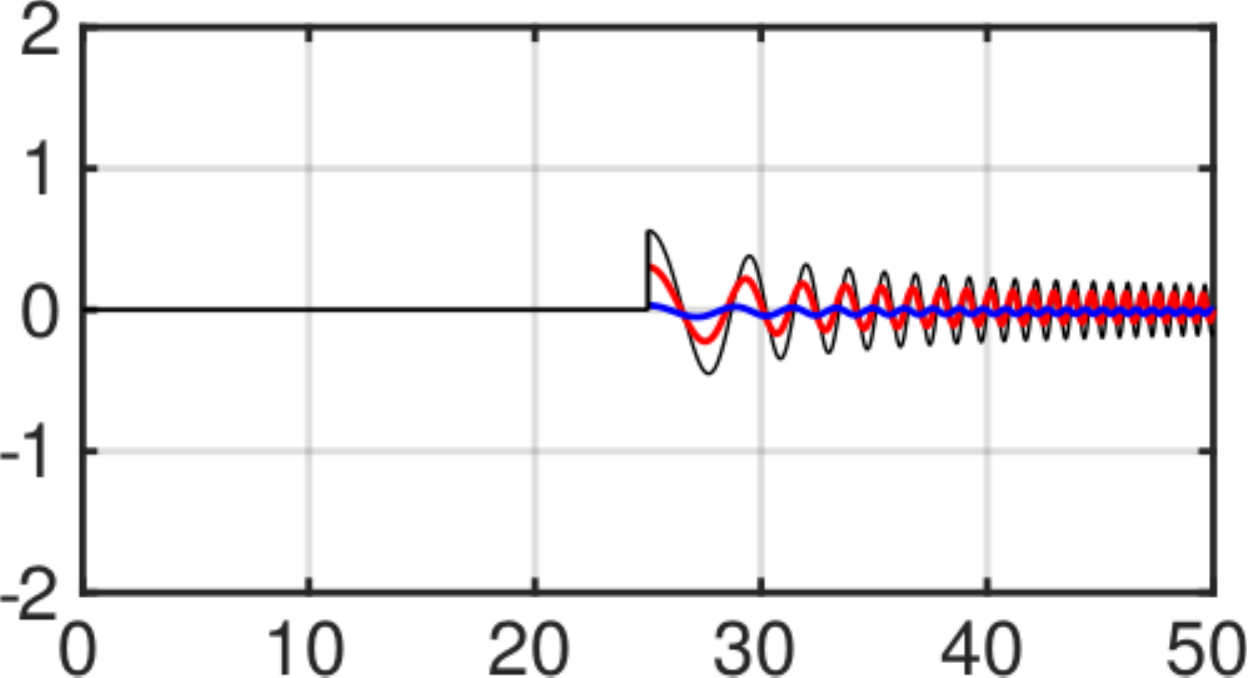}}
{$ $}{-3mm}{\begin{rotate}{90}$v_{x}^{(w)}$\end{rotate}}{1mm}
\end{minipage}
\begin{minipage}[c]{.26\linewidth}
\FigureXYLabel{\includegraphics[width = 1.\linewidth]{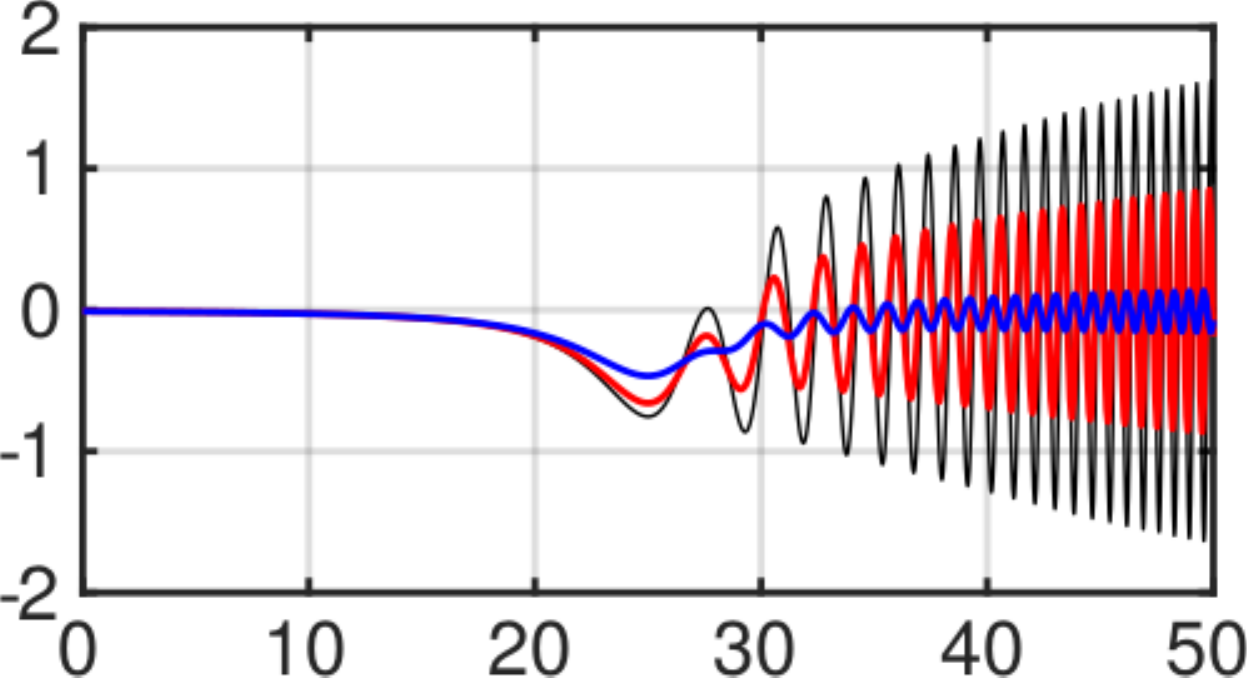}}
{$ $}{-3mm}{\begin{rotate}{90} {\small $v_y$} \end{rotate}}{1mm}
\end{minipage}
\hspace{0.5cm}
\begin{minipage}[c]{.26\linewidth}
\FigureXYLabel{\includegraphics[width = 1.\linewidth]{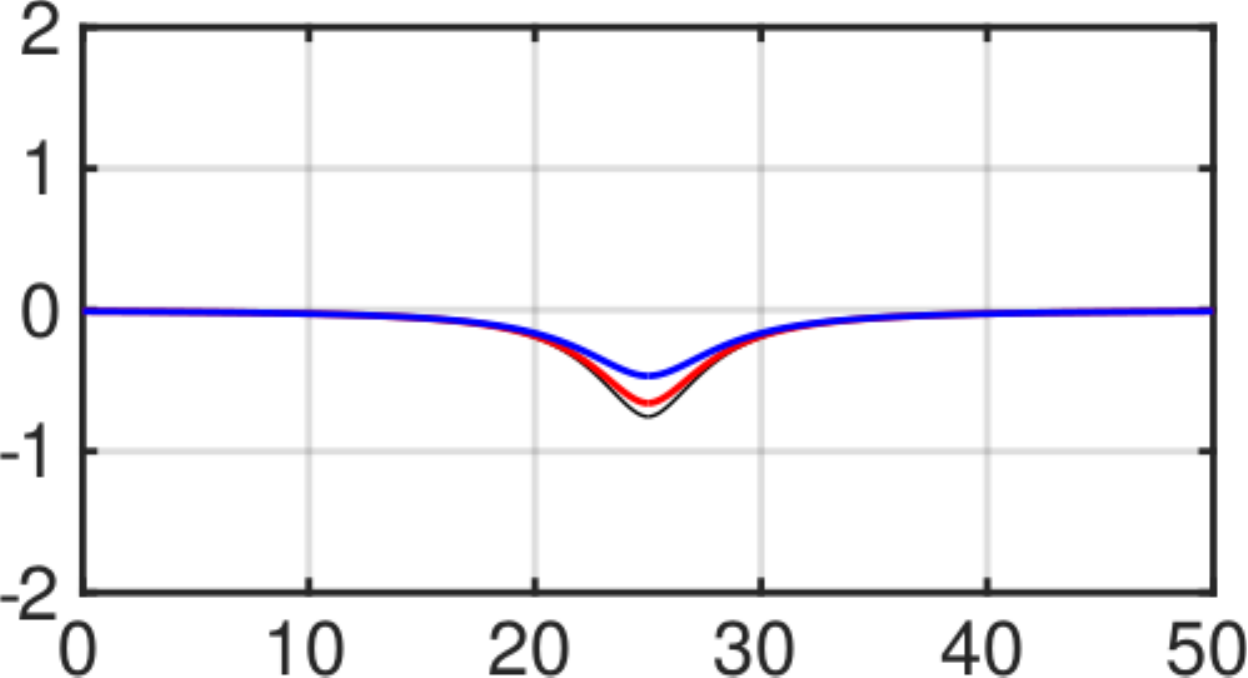}}
{$ $}{-1mm}{\begin{rotate}{90}$v_y^{(v)}$\end{rotate}}{1mm}
\end{minipage}
\hspace{0.5cm}
\begin{minipage}[c]{.26\linewidth}
\FigureXYLabel{\includegraphics[width = 1.\linewidth]{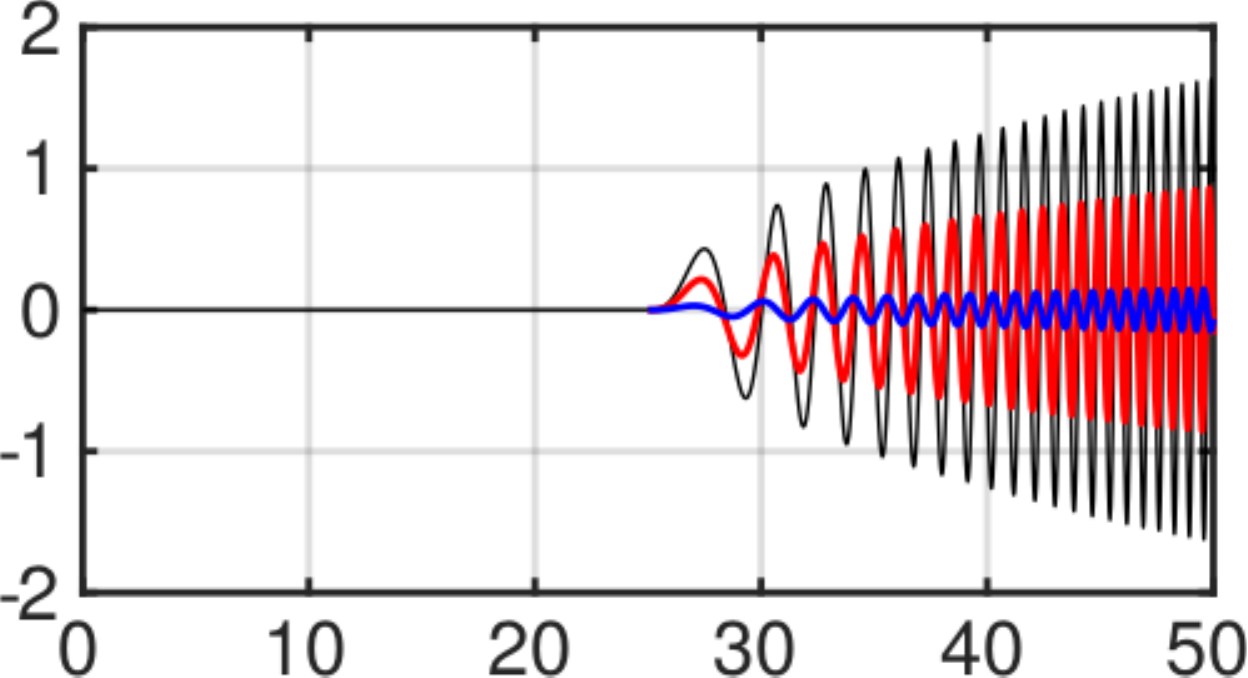}}
{$ $}{-1mm}{\begin{rotate}{90}$v_y^{(w)}$\end{rotate}}{1mm}
\end{minipage}
\hspace{0.5cm}
\begin{minipage}[c]{.26\linewidth}
\FigureXYLabel{\includegraphics[width = 1.\linewidth]{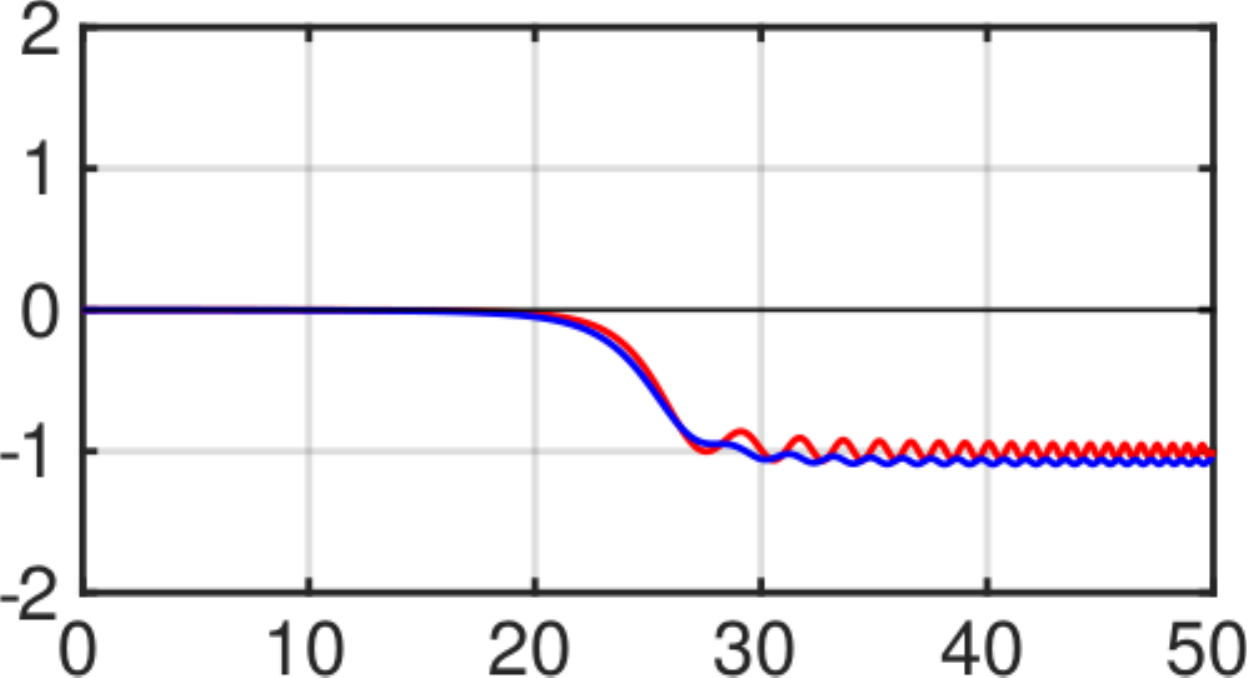}}
{$ $}{-3mm}{\begin{rotate}{90} {\small $v_{z}$}\end{rotate}}{1mm}
\end{minipage}
\hspace{0.5cm}
\begin{minipage}[c]{.26\linewidth}
\FigureXYLabel{\includegraphics[width = 1.\linewidth]{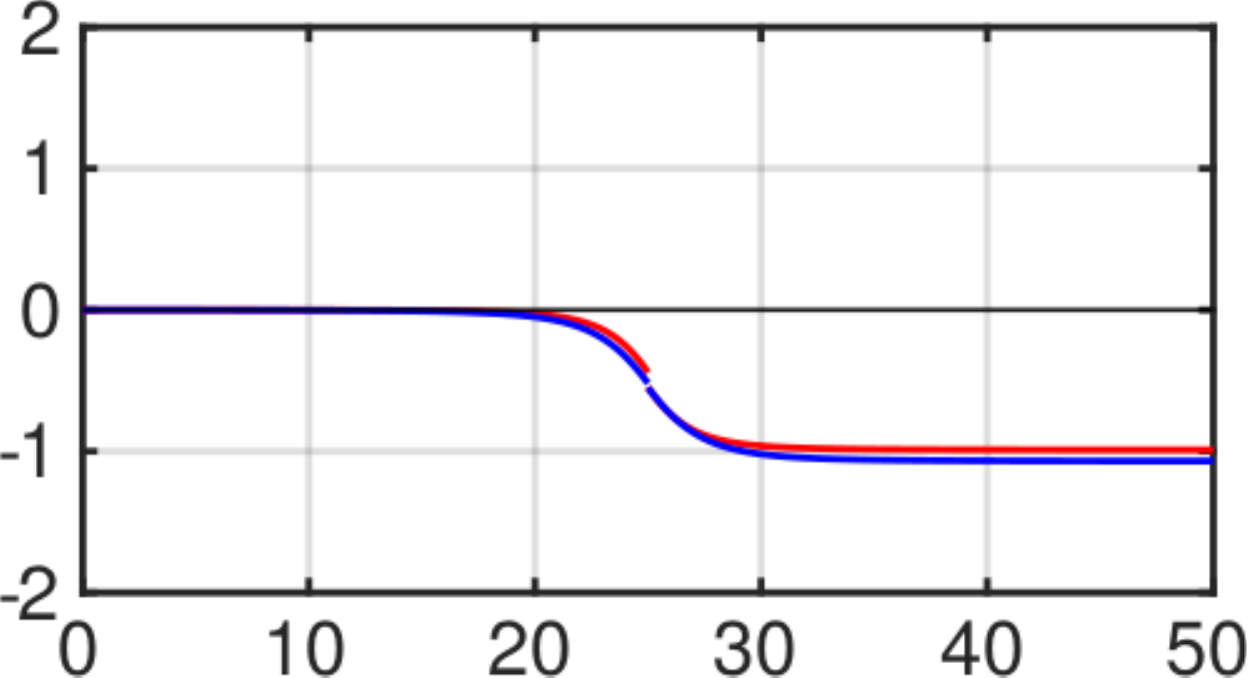}}
{$ $}{-3mm}{\begin{rotate}{90}$v_{z}^{(v)}$\end{rotate}}{1mm}
\end{minipage}
\hspace{0.5cm}
\begin{minipage}[c]{.26\linewidth}
\FigureXYLabel{\includegraphics[width = 1.\linewidth]{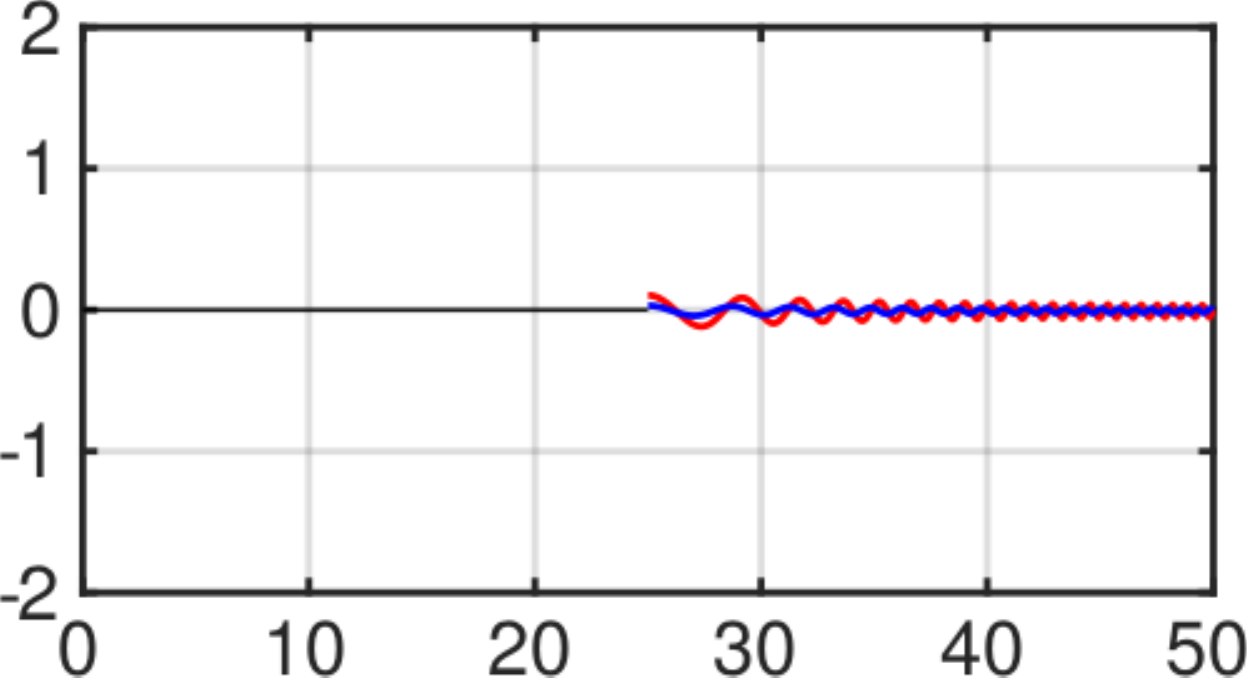}}
{$ $}{-3mm}{\begin{rotate}{90}$v_{z}^{(w)}$\end{rotate}}{1mm}
\end{minipage}
\hspace{0.5cm}
\begin{minipage}[c]{.26\linewidth}
\FigureXYLabel{\includegraphics[width = 1.\linewidth]{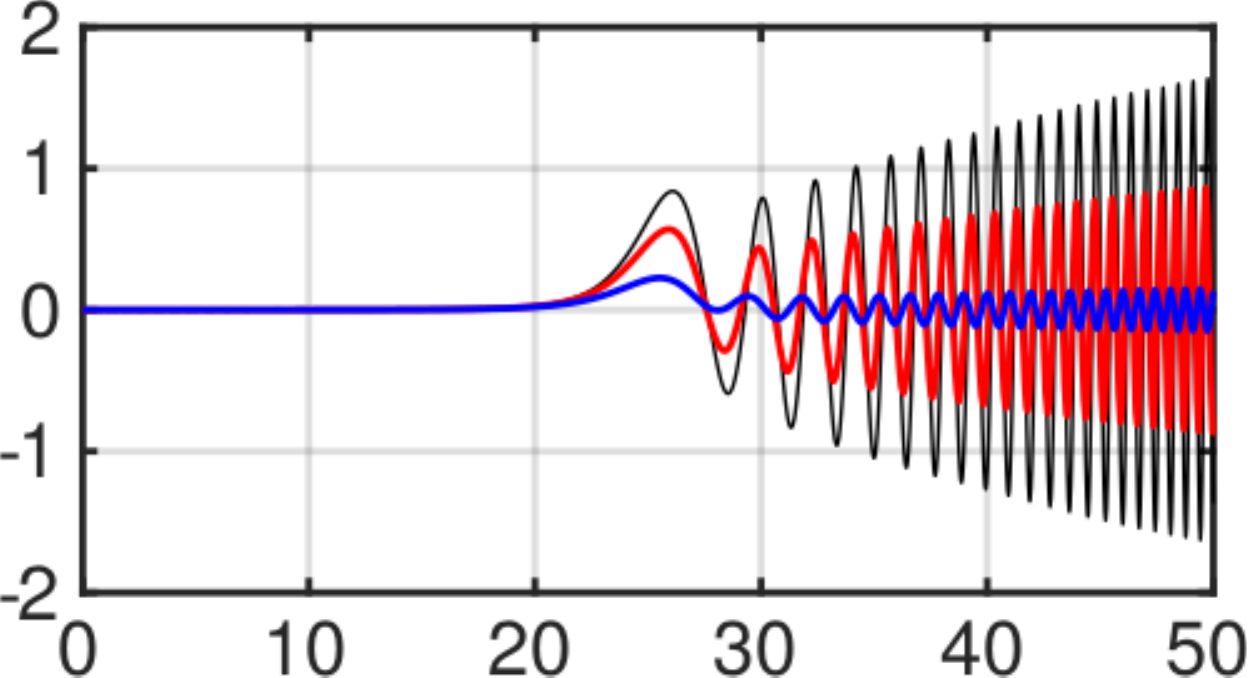}}
{$\tau$}{-3mm}{\begin{rotate}{90}{\small $D$}\end{rotate}}{1mm}
\end{minipage}
\hspace{0.5cm}
\begin{minipage}[c]{.26\linewidth}
\FigureXYLabel{\includegraphics[width = 1.\linewidth]{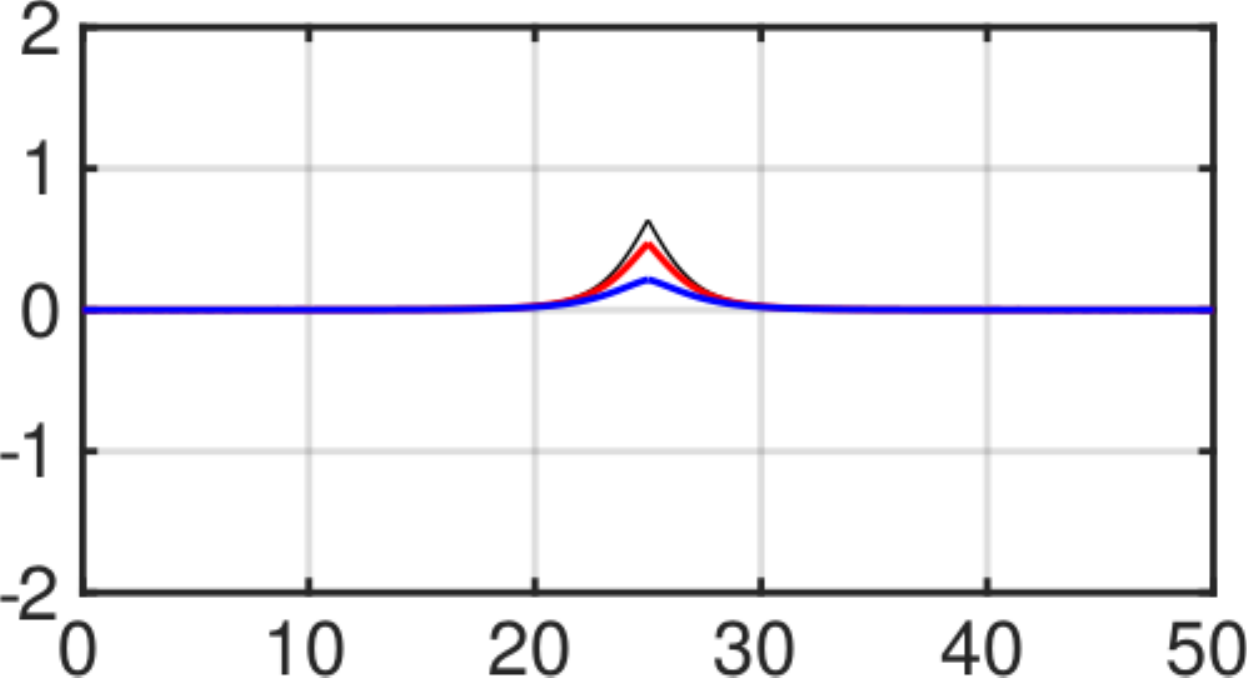}}
{$\tau$}{-3mm}{\begin{rotate}{90}$D^{(v)}$\end{rotate}}{1mm}
\end{minipage}
\hspace{0.5cm}
\begin{minipage}[c]{.26\linewidth}
\FigureXYLabel{\includegraphics[width = 1.\linewidth]{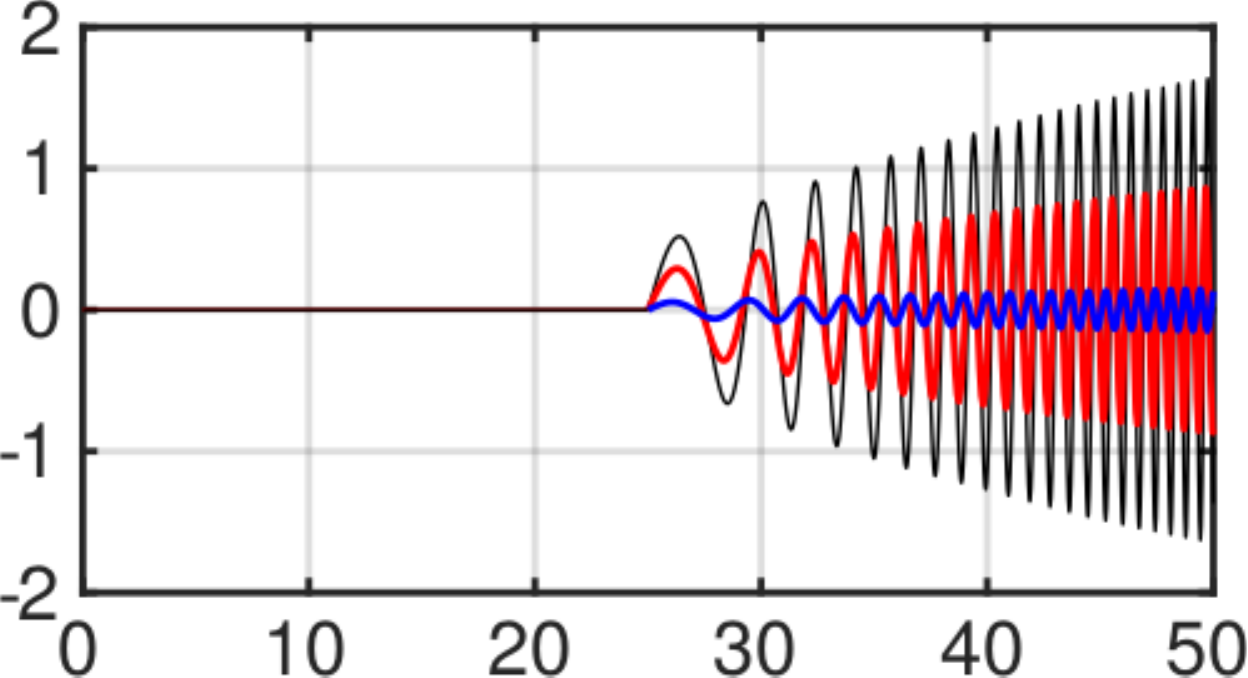}}
{$\tau$}{-3mm}{\begin{rotate}{90}$D^{(w)}$\end{rotate}}{1mm}
\end{minipage}
\caption{The dynamics of $(v_x,v_y,v_z,D)$ (left column) and their
vortex (v) (central column) and wave (w) (right column) parts  for
the initially pure vortex harmonic
($v_x^\mathrm{(w)}(0),v_y^\mathrm{(w)}(0),v_z^\mathrm{(w)}(0),D^\mathrm{(w)}(0)
= 0$) for $\mathcal{M}=0.4$, $\beta(0)=10$ and $\gamma=0,~0.5,~1$.
Black, red and blue lines correspond to 2D ($\gamma=0$), 3D
($\gamma=0.5$) and ($\gamma=1$) cases, respectively. For the
considered Mach number the wave generation is maximal for 2D case.
For $\gamma=0.5$ case, the wave generation is somewhat reduced
compared to $\gamma=0$ case. For $\gamma=1$ case, the wave
generation is substantially reduced (almost negligible) compared to $\gamma=0$ case.}
\label{lin_case}
\end{figure}

The plots in the left column in Figure \ref{lin_case} show the
dynamics of $v_x$, $v_y$, $v_z$ and $D$ at ${\mathcal {M}}=0.4$,
$\beta(0) = 10$, $\gamma = 0,~0.5,~ 1$. The black, red and blue
curves represent time evolution of harmonics with $\gamma =
0,~0.5,~1$, respectively. The imposed vortex mode harmonic remains
aperiodic up to $\tau < \beta(0)/\mathcal{M} \equiv \tau^{\ast}=25$,
i.e., at $\beta(\tau)>0$. The oscillating/wave nature appears just
at $\tau>\tau^{\ast}$, i.e., at $\beta(\tau)<0$. The shear-wise wave
number of the harmonics, $\beta(\tau)$ (or $k_y(t)$ in dimensional
units), is time dependent. One can say, that the harmonics ``drift''
in spectral space along the $k_y$ axis and each wave harmonic
emerges from the related vortex one when crossing of the $k_x$
axis (or $k_y=0$ line) during this ``drift'', that occurs at the
critical time $\tau^\ast$ that is a universal timescale for the
appearance of linearly generated waves.

This peculiar behavior motivates the introduction of a numerical
mode-splitting at $\tau=\tau^{\ast}$, in spite of the strong
vortex--wave mode coupling in the its vicinity \cite{Chagelishvili1997}. 
Although the wave
and vortex mode parts are not distinguishable at
$\vert\beta(\tau)\vert<1$, the mathematical separation is useful
as it allows to understand the different behavior of the vortex and
wave mode harmonics during their further dynamics. It is possible to
extract a vortex mode harmonic at $\tau=\tau^{\ast}_+$ (where
$\tau^\ast_{\pm} \equiv \tau^\ast \pm 0$), that further evolves
independently of the wave mode harmonic and retains its aperiodic
nature. We can therefore split the velocity and density perturbations as
\begin{align}
v_x(\tau^{\ast}_-) &\equiv v^\mathrm{(v)}_x(\tau^{\ast}_-) =
v^\mathrm{(v)}_x(\tau^{\ast}_+) + v^\mathrm{(w)}_x(\tau^{\ast}_+), \nonumber \\
v_y(\tau^{\ast}_-) &\equiv v^\mathrm{(v)}_y(\tau^{\ast}_-) =
v^\mathrm{(v)}_y(\tau^{\ast}_+) + v^\mathrm{(w)}_y(\tau^{\ast}_+), \nonumber \\
v_z(\tau^{\ast}_-) &\equiv v^\mathrm{(v)}_z(\tau^{\ast}_-) =
v^\mathrm{(v)}_z(\tau^{\ast}_+) + v^\mathrm{(w)}_z(\tau^{\ast}_+), \nonumber \\
D(\tau^{\ast}_-)   &\equiv D^\mathrm{(v)}(\tau^{\ast}_-) =
D^\mathrm{(v)} (\tau^{\ast}_+) + D^\mathrm{(w)} (\tau^{\ast}_+), \nonumber
\end{align}
where the superscripts $\mathrm{(v)}$ and $\mathrm{(w)}$ denote the
vortex and wave parts, respectively. The plots in the central column
of Figure \ref{lin_case} show the dynamics of the vortex part of the related
quantities whereas the plots in the right column show the dynamics of the
wave part.
To summarize the results of the splitting:
\\
\emph{(i)} There is an abrupt emergence of a acoustic wave harmonic
from the vortex one at the crossing of the $k_x$-axis during the ``drift'';\\
\emph{(ii)} The generation mechanism is anisotropic -- the generated
acoustic wave harmonics can be found only in those parts of spectral space where $k_y/k_x \leq 0$;\\
\emph{(iii)} The gap in the evolution of the vortical part of the streamwise velocity perturbation,
$v^\mathrm{(v)}_x(\tau^{\ast}_+) - v^\mathrm{(v)}_x(\tau^{\ast}_-)$, is a measure of
the spontaneously generated wave harmonic;\\
\emph{(iv)} The wave harmonics emerge abruptly with very regular
phase. At the moment of the emergence the shearwise velocity and
density of the wave harmonic
($v^\mathrm{(w)}_y(\tau^{\ast}_+),~D^\mathrm{(w)}
(\tau^{\ast}_+)=0$) are zero, while the values of the streamwise and
spanwise velocities
($v^\mathrm{(w)}_x(\tau^{\ast}_+),~v^\mathrm{(w)}_z(\tau^{\ast}_+)$)
are maximal;\\
\emph{(v)} The mutual Independence of the dynamics of the vortex and wave parts
of the quantities occurs for $\tau > \tau^{\ast}$ indicating
the validity of the splitting at $\tau = \tau^{\ast}$;\\
\emph{(vi)} For the considered mode-dependent Mach number,
${\mathcal {M}}=0.4$, the generation of 2D wave harmonics prevails
over 3D ones -- the generation of 3D wave harmonics decreases
rapidly with increasing $\gamma=k_z/k_x$.

We should note additionally, that the topology of the summarized
anisotropic linear generation mechanism differs principally from the
topology of the linear source representations in the classical
formulations of the acoustic analogy \cite{Hau2015}. The linear
source terms of the latter are distributed in all regions/quadrants
of the spectral space ($k_y/k_x>0$ and $k_y/k_x<0$) predicting the
immediate emergence of wave harmonics in any of the quadrants. This
makes acoustic analogies incompatible with the linear mechanism of
generation of wave harmonics by vortex mode harmonics in shear flows,
induced by the non-normality.

According to the literature, an entire wave-packet is the
source of the acoustic wave generation and the latter process is
described in the framework of the modal/spectral approach. However,
in fact, wave-packets are composed of coherent vortex mode harmonics
that represent the basic ``elements'' of dynamic processes under
linear shear \cite{Yoshida2005}. There is an abrupt emergence of
waves from the initial vortex mode at $\tau = \tau^\ast$ with the
amplitude of streamwise and spanwise velocities,
$v_x^\mathrm{(w)}(\tau^\ast) =
v_x^\mathrm{(v)}(\tau^\ast_-)-v_x^\mathrm{(v)}(\tau^\ast_+)$ and
$v_z^\mathrm{(w)}(\tau^\ast) =
v_z^\mathrm{(v)}(\tau^\ast_-)-v_z^\mathrm{(v)}(\tau^\ast_+)$.
Consequently, the basic process is the linear generation of acoustic
wave harmonics by the related vortex modes. Therefore, the basic source of
acoustic waves are vortex mode harmonics and not an entire/undivided
wave-packet -- the packet just collects via constructive
interference the acoustic wave harmonics generated by the basic
process.

\subsection*{Efficacy of the wave generation}
The efficacy of the wave generation process is identified by the
jump occurring in the vortex part of the dependent (split) variables
$(v_x^\mathrm{(v)},~v_y^\mathrm{(v)},~v_z^\mathrm{(v)},D^\mathrm{(v)})$
at $\tau = \tau^\ast$ and the thereby generated wave harmonic 
\cite{Chagelishvili2002a,Tevzadze2006}. This
is directly seen in Figure \ref{lin_case} for the streamwise and
spanwise perturbation velocities. Although not visible, the jump
naturally occurs in the derivatives of the remaining quantities.
Thus, we define the efficacy of the wave generation process by 
initially pure vortex mode perturbations as the ratio of the 
wave energy to the vortex one
\begin{equation}
 \label{effic}
\eta \equiv\frac{E^\mathrm{(w)}(\tau^\ast_+)}{E^\mathrm{(v)}(\tau^\ast_-)}=
\frac{\vert v_x^{\mathrm{(w)}}(\tau^\ast_+)\vert^2 + \vert v_z^{\mathrm{(w)}}(\tau^\ast_+)\vert^2}
{\vert v_x^{\mathrm{(v)}}(\tau^\ast_-)\vert^2 + \vert v_y^{\mathrm{(v)}}(\tau^\ast_-)\vert^2 +
\vert v_z^{\mathrm{(v)}}(\tau^\ast_-)\vert^2 + \vert D^{\mathrm{(v)}}(\tau^\ast_-)\vert^2}~,
\end{equation}
where the fact that $\vert v_y^{\mathrm{(w)}}(\tau^\ast_+)\vert^2=0$ and
$\vert D^{\mathrm{(w)}}(\tau^\ast_+)\vert^2 =0$ is taken into account.
\begin{figure}[H]
\centering
\begin{minipage}[c]{.53\linewidth}
\FigureXYLabel{\includegraphics[width=.9\textwidth]{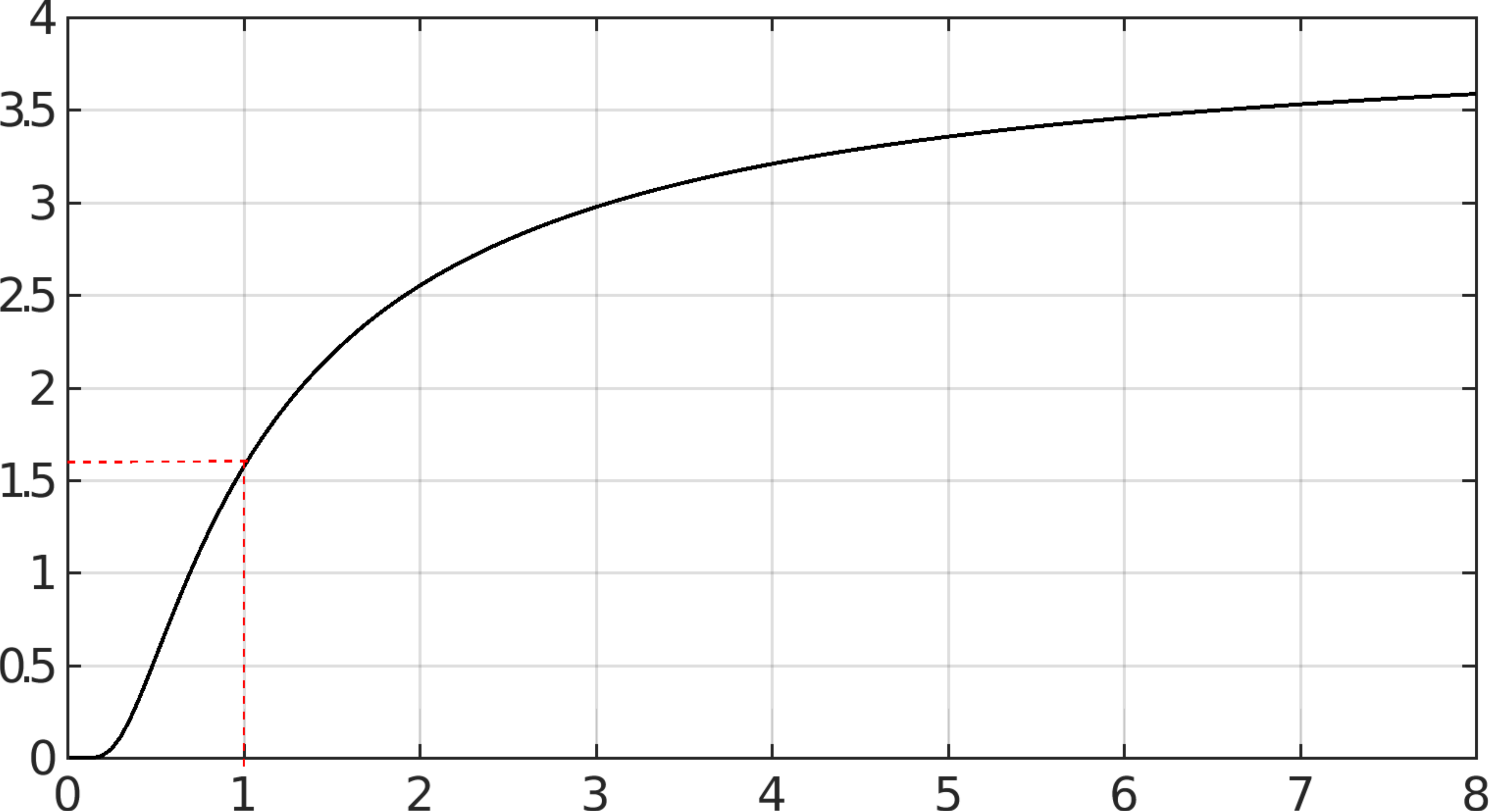}}
{$\mathcal{M}$}{-1mm}{\begin{rotate}{90}$\eta$\end{rotate}}{2mm}
\end{minipage}
\hspace{0.2cm}
\begin{minipage}[c]{.2\linewidth}
\FigureXYLabel{\includegraphics[width=.895\textwidth]{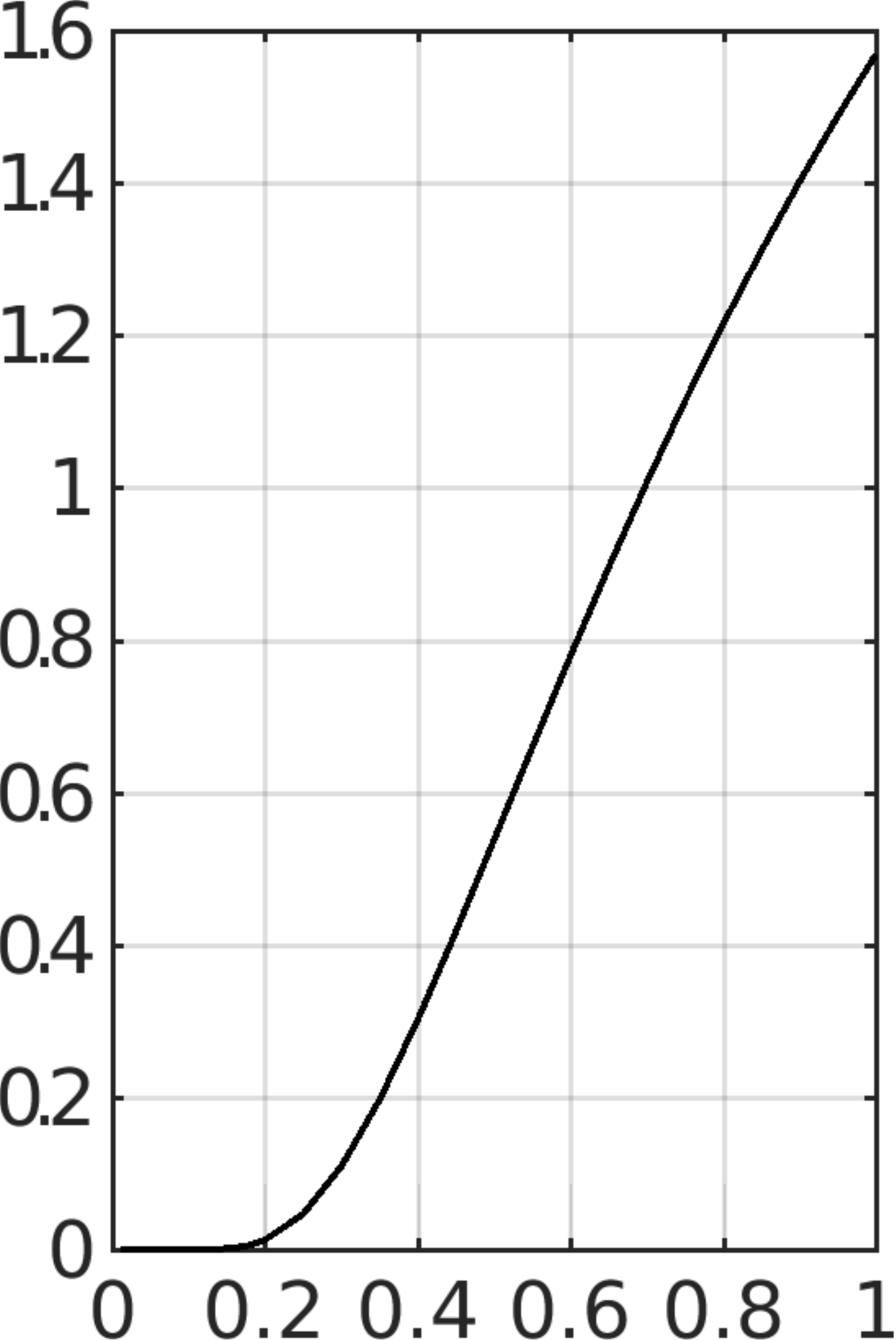}}
{$\mathcal{M}$}{-1mm}{\begin{rotate}{90}$\eta$\end{rotate}}{2mm}
\end{minipage}
\caption{The dependence of the effective parameter of linear
generation of acoustic wave harmonics by related vortex mode ones,
$\eta$, on $\mathcal{M}$ for the 2D case (i.e., at $\gamma=0$) for
$\beta(0) = k_y(0)/k_x=10$. The left plot is for a wide range of
Mach number values, while the right plot shows the zoomed view of
$\eta$ for $\mathcal{M} \in [0,1]$.}
\label{fig:eta}
\end{figure}
In Figure \ref{fig:eta}  the dependence of the effective parameter
of wave generation, $\eta$, on $\mathcal{M}$ for the 2D case (i.e., at
$\gamma=0$) is presented. The left plot is for a wide range of Mach
number values, while the right one shows a zoomed view of $\eta$
for the region $\mathcal{M} \in [0,1]$.
The process of the wave generation becomes noticeable at
$\mathcal{M}=0.2$ and substantial already at $\mathcal{M}=0.3$. The
growth of $\eta$ becomes gentle from $\mathcal{M}=2$ and
asymptotically tends to $4$ at $\mathcal{M} \rightarrow \infty$. The
right plot clearly shows a sharp change of $\eta$ for a moderate
$\mathcal{M} \in [0.3,1]$. We operate in the range of $\mathcal{M} \in [0.1, 1]$
mode-dependent Mach numbers to establish the importance of the
discussed linear mechanism of wave generation in the case of a
compressible turbulent time-developing mixing layer, the results of that numerical
study is presented in the next section. One has to note, that $\eta$ is not
the efficiency of the wave generation itself, since it reaches values larger
than one. This indicates that the wave energy growth is provided by the flow shear,
while vortex disturbances only trigger/mediate the generation process.

Equation \eqref{effic} for the 3D case has to be evaluated
numerically, as the jump occurring around $\tau = \tau^\ast$ does
not follow an exact analytical relation contrary to the 2D case.
Hereby, we fix the initial wavenumber ratio $\beta(0) = k_y(0)/k_x=10$, while
varying the perturbation Mach number $\mathcal{M}$ and the
wavenumber ratio $\gamma = k_z/k_x$. From figure \ref{fig:eta} it follows that the
efficacy of the wave generation, specifically, for
$\mathcal{M}=0.4$ and 2D harmonics is $\eta(\mathcal{M}=0.4,\gamma=0)=0.32$. The calculations
show that the efficacy decreases with $\gamma$:
$\eta(\mathcal{M}=0.4,\gamma = 0.5)=0.15$, $\eta(\mathcal{M}=0.4,\gamma=1)=0.0022$ and
$\eta(\mathcal{M}=0.4,\gamma=2)=3.2\cdot10^{-7}$, etc.

Generally, the efficacy of the linear mechanism of wave generation by a pure
vortex mode harmonic for moderate mode-dependent Mach numbers,
$\mathcal{M} \lesssim 1$, reaches maximum for 2D harmonics and
decreases quite fast with an increase of $\gamma$. Such
behavior of $\eta(\mathcal{M},\gamma)$ at $\mathcal{M} \in [0.1,1]$
is a guideline for establishing the basic mechanism of
acoustic wave generation in a compressible turbulent time-developing
mixing layer, studied numerically in the next section.

\section{DNS of compressible time-developing mixing layer}
\label{sec:DNS}
The simulations are performed to capture the essence -- the basic
mechanism -- of the acoustic wave generation in shear flows by means of a compressible
turbulent time-developing mixing layer. We solved the Navier-Stokes
equations numerically using the optimized summation-by-parts
dispersion relation preserving finite difference scheme of
sixth-order accuracy in space and a third-order low-storage
Runge-Kutta scheme in time \cite{Pantano2002,Foysi2009}. The
convective Mach number of the time-developing shear layer between
two ($U_1$ and $U_2$) anti-parallel streams of fluid with equal absolute
values of velocities, $U_1=-U_2=\Delta  U/2$, and constant density
and sound speed, is defined by 
\begin{equation}
\label{Mc}
M_c = \frac{\Delta U}{2c_s}.
\end{equation}
\begin{figure}
\centering
\begin{minipage}[c]{.4\linewidth}
\FigureXYLabel{\includegraphics[width = 1.\linewidth]{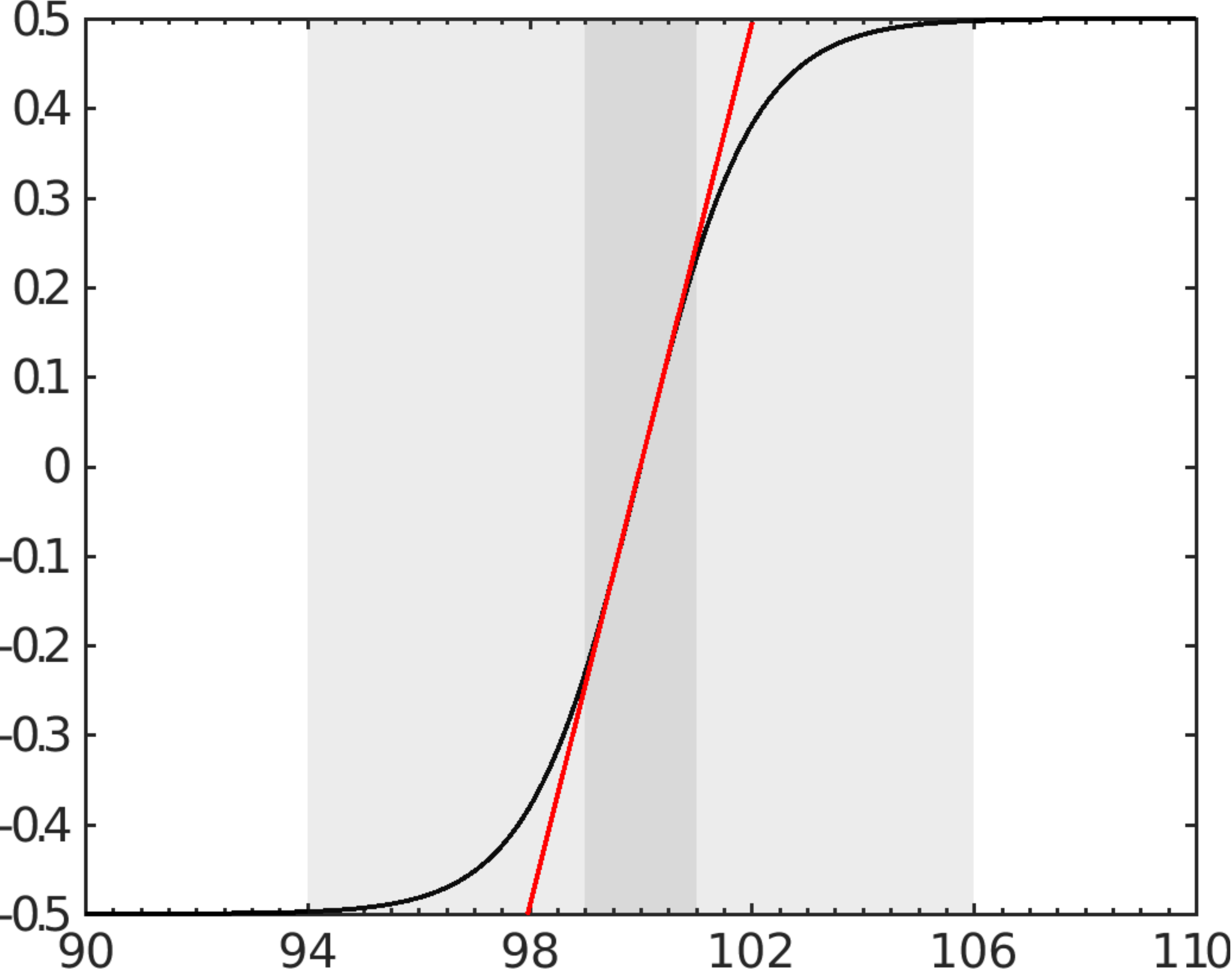}}
{$y$}{1mm}{\begin{rotate}{90}$\overline{U}$\end{rotate}}{3mm}
\end{minipage}
\caption{Mean velocity profile at initial moment of time. The red
line shows the maximum gradient of the mean velocity profile. One
can consider the grey area as the shear layer area (where the mean
velocity gradient is non-zero). The dark grey area represents the
central/body/core region of the flow with linear profile of mean
velocity which matches to the red line.}
\label{Um_t0}
\end{figure}
The initial velocity profile is given as a hyperbolic tangent function
for the mean streamwise velocity and shearwise and spanwise mean velocity
components are set to zero (see Figure \ref{Um_t0}):
\begin{equation}
\label{U(y)}
U(y)= \frac{\Delta U}{2}\tanh \left (\frac{y}{2{\delta}_{\theta}(0)} \right ),
~ V = 0, ~ W = 0,
\end{equation}
where ${\delta}_{\theta}(0)$ is the initial momentum thickness of
shear layer. The momentum thickness of shear layer,
${\delta}_{\theta}(t)$, varies in time and represent s one 
of appropriate length scales used to
describe the flow. The vorticity thickness, $\delta_\omega(t)$, is
an another widely used length scale. The definitions have the
form:
\begin{equation}
\label{Reynolds_numbers}
\delta_\theta (t) = \frac{1}{\rho_0}\int\limits_{-L_y/2}^{L_y/2} {\bar \rho}\left[ \frac{1}{4} -\left( \frac{{\bar u_1}}{\Delta U}\right )^2  \right ]dy; ~~~
\delta_\omega(t) = \frac{\Delta U}{|dU/dy|_{max}}; ~~~
 Re_{\delta_{\theta}(t)}=\frac{\Delta U \delta_{\theta }(t)}{\nu};  ~~~ Re_{\delta_{\omega}(t)}=\frac{\Delta U \delta_{\omega }(t)}{\nu}.
\end{equation}
Broadband disturbances are superimposed on the basic flow to trigger
transition to turbulence in the flow \cite{Foysi2009}. The initial
extent of the superimposed perturbations is limited in the shearwise
direction to the initial shear layer thickness (the grey area on
Figure \ref{Um_t0}). The turbulence develops in time leading to an
increase of the shear layer thickness.

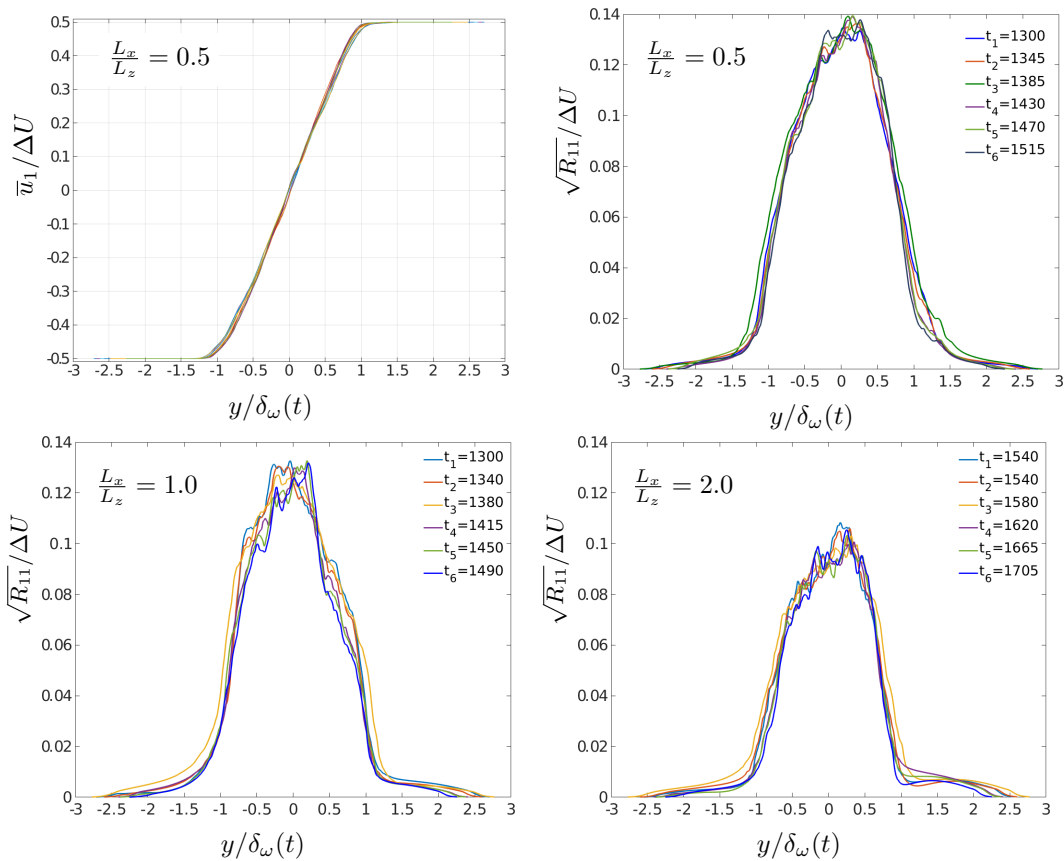
\begin{figure}[H]
\centering
\begin{minipage}[c]{.38\linewidth}
\FigureXYLabel{\includegraphics[width = .9\linewidth]{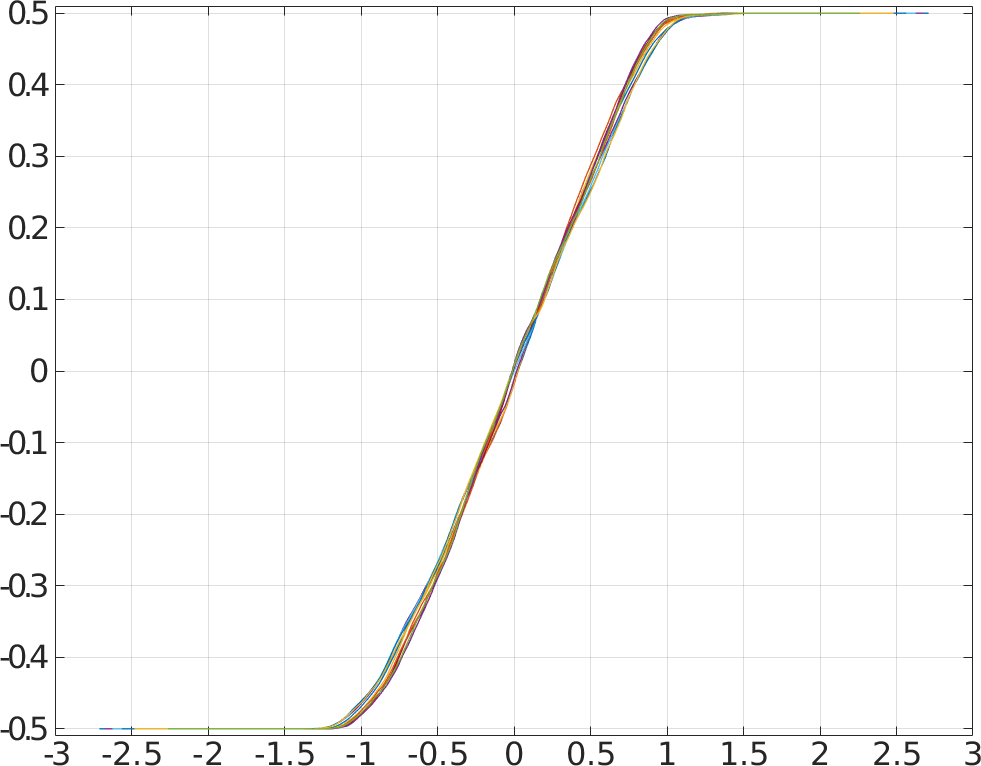}}
{$y/\delta_{\omega}(t)$}{1mm}{\begin{rotate}{90}${\overline u}_1/\Delta U$\end{rotate}}{2mm}
\end{minipage}
\begin{tikzpicture}[overlay]
\fill[fill=white]
(-5,2.2) node[fill=white] {$\frac{L_x}{L_z}=0.5$};
 \end{tikzpicture}
 \hspace{0.3cm}
\begin{minipage}[c]{.35\linewidth}
\FigureXYLabel{\includegraphics[width = 1.\linewidth]{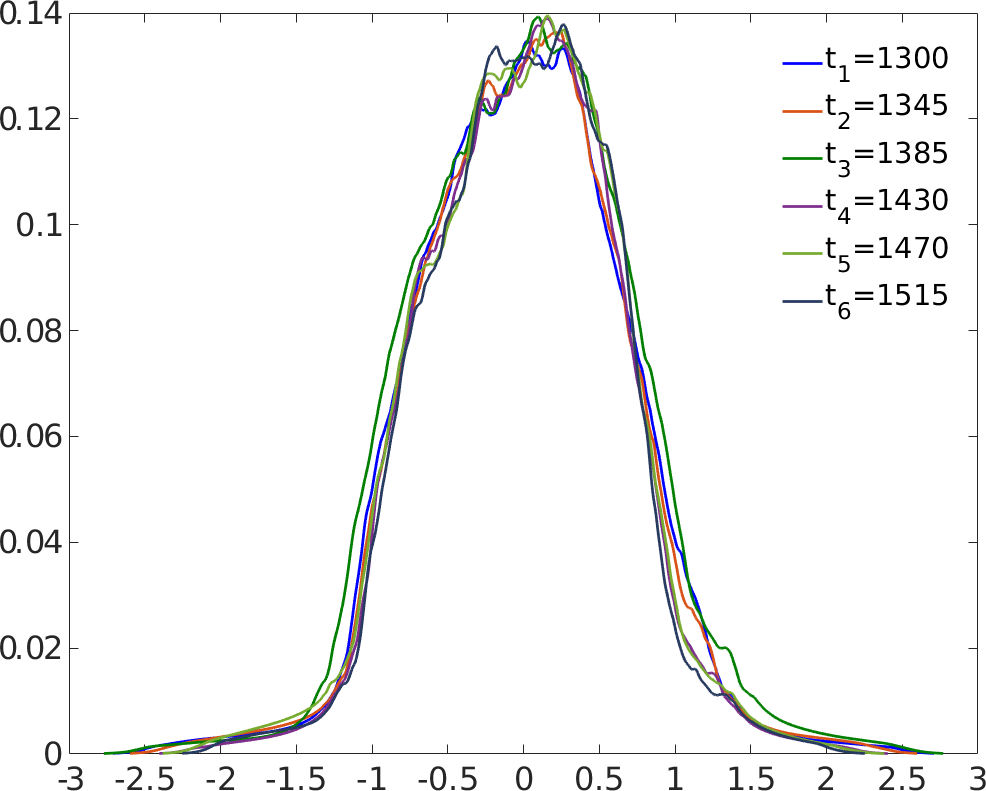}}
{$y/\delta_{\omega}(t)$}{1mm}{\begin{rotate}{90}\small $\sqrt{R_{11}}/\Delta U$\end{rotate}}{2mm}
\end{minipage}
\begin{tikzpicture}[overlay]
\fill[fill=white]
(-4.8,2.2) node[fill=white] {$\frac{L_x}{L_z}=0.5$};
 \end{tikzpicture}
\hfill
\begin{minipage}[c]{.35\linewidth}
\FigureXYLabel{\includegraphics[width = 1.\linewidth]{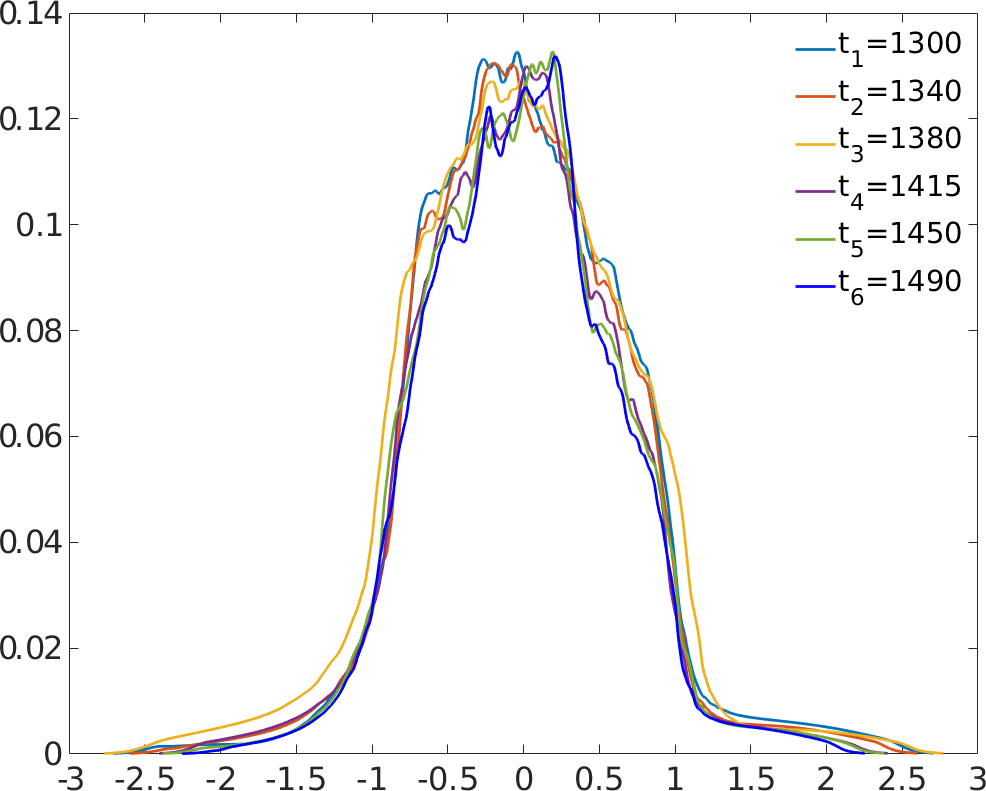}}
{$y/\delta_{\omega}(t)$}{1mm}{\begin{rotate}{90}\small $\sqrt{R_{11}}/\Delta U$\end{rotate}}{2mm}
\end{minipage}
\begin{tikzpicture}[overlay]
\fill[fill=white]
(-4.8,2.2) node[fill=white] {$\frac{L_x}{L_z}=1.0$};
 \end{tikzpicture}
\hspace{0.5cm}
\begin{minipage}[c]{.35\linewidth}
\FigureXYLabel{\includegraphics[width = 1.\linewidth]{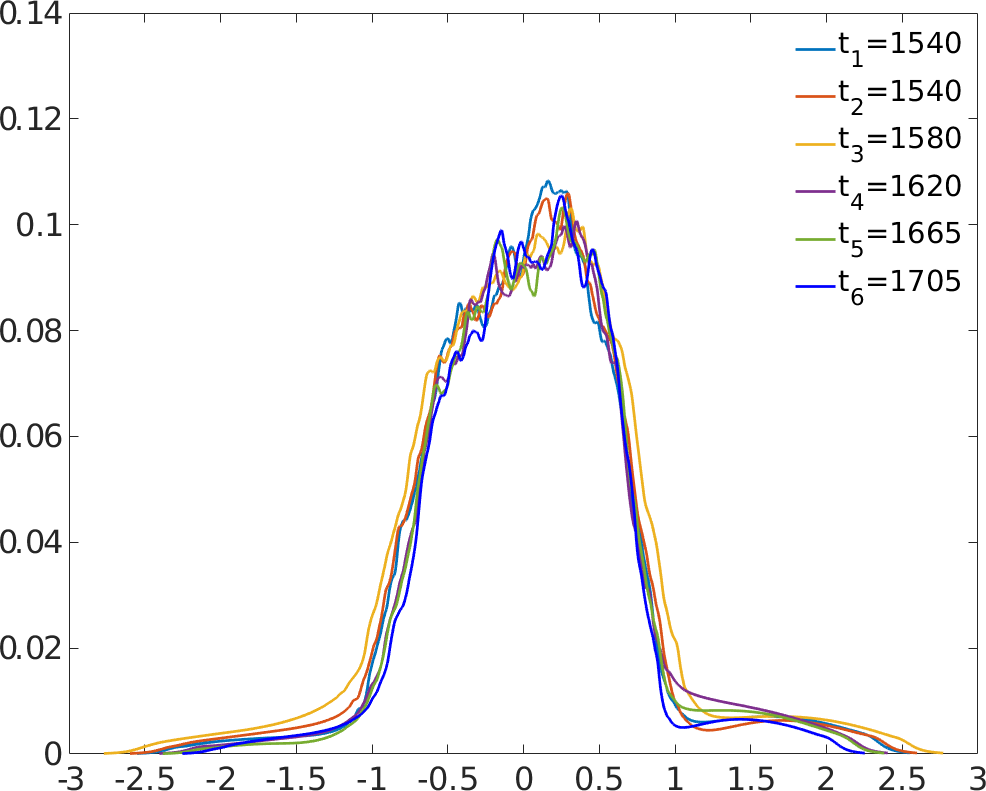}}
{$y/\delta_{\omega}(t)$}{1mm}{\begin{rotate}{90}\small $\sqrt{R_{11}}/\Delta U$\end{rotate}}{2mm}
\end{minipage}
\begin{tikzpicture}[overlay]
\fill[fill=white]
(-4.8,2.2) node[fill=white] {$\frac{L_x}{L_z}=2.0$};
 \end{tikzpicture}
\caption{Streamwise velocity profiles for $L_x/L_z=0.5$ in the
self-similar period of development of the flow; RMSs of streamwise
velocity in self-similar period of development at multiple time
samples are presented. From left to right: $L_x/L_z = 0.5$;
$L_x/L_z = 1.0$, $L_x/L_z = 2.0$. }
\label{ss_R11}
\end{figure}

The numerical parameters of the numerical simulations presented are given in Table \ref{DNSparametersMain}. 
We performed simulations for 
the following convective Mach numbers $M_c = 0.3,~0.7$. The initial Reynolds numbers based on the
momentum and vorticity thicknesses  at the initial moment of time were chosen as $Re_{\delta _{\theta }(0)} = 800$
and $Re_{\delta _{\omega }(0)} = 3200$. The last column of  the table represents the growth rate of the  
shear layer in the self-similar period of evolution in time  when the linear growth rate is achieved. 
\begin{table}[H]
\vspace{0mm}
\small
\begin{center}
\begin{tabular}{c c c c c c c c c}
{\small Case} ~&{\small $({L_x \times L_y \times L_z})/{\delta_{\theta}(0)}$} & {\small $N_x\times N_y \times N_z$} ~& {\small ${L_x}/{L_z}$} ~& $M_c$~& ~ {\small  $Re_{\delta_{\theta}(0)}$}  ~& ~ {\small $Re_{\delta_{\theta}}$} ~ & {\small $Re_{\delta_{\omega}}$} ~ & ~{\small $\dot{\delta_\theta}$}\\
\hline \hline
$1$  &$100 \times 200 \times 50$   & $1024 \times 1025 \times 512$    & $2.0$ & ~$0.7$  & $800$ & $9256$   & $ 37 040$ & ~~$0.0047 $  \\
$2$  &$100 \times 200 \times 100$ & $1024 \times 1025 \times 1024$ ~& $1.0$ & ~$0.7$ & $800$ & $11256$ &  $43 920$ & ~~ $0.0062 $ \\
$3$  &$100 \times 200 \times 200$ & $1024 \times 1025 \times 2048$ & $0.5$ &~$0.7$ &$800$  & $12593$ &   $47414$ & ~~$0.0072 $\\
\hline
$4$ & $100 \times 200 \times 50$   & $512  \times 1025 \times 256$    & $2.0$ & $0.3$  & $800$ & $7 230$   & $ 27 500$ & ~ $0.0081$ \\
$5$ &~$100 \times 200 \times 100$ & $512 \times 1025 \times 512$   & $1.0$ & $0.3$ & $800$ & $7 472$ &  $27 970$ & ~  $0.0106$\\
$6$ &$100 \times 200 \times 200$ & $512 \times 1025 \times 1024$ & $0.5$ &$0.3$ &$800$  & $6 211$ &  $29 736$  &~ $0.0120$ \\
\end{tabular}
\caption{Simulation parameters: $L_i$ and $N_i$ ($i=x,y,z$) are
domain lengths  and number of grid points correspondingly. The
simulations were performed for the convective Mach numbers  $M_c =
0.3,~0.7$. $Re_{\delta_{\theta}}$ and $Re_{\omega}$ are final
Reynolds numbers based on the momentum and vorticity thicknesses respectively. 
${\dot {\delta}_{\theta}} = {d(\delta_\theta({t}))}/{d{t}}$ is the growth rate of 
shear layer within the self-similar period of evolution in time.} \label{DNSparametersMain}
\end{center}
\end{table}
\begin{figure}[H]
\centering
\begin{minipage}[c]{.43\linewidth}
\FigureXYLabel{\includegraphics[width = .9\linewidth]{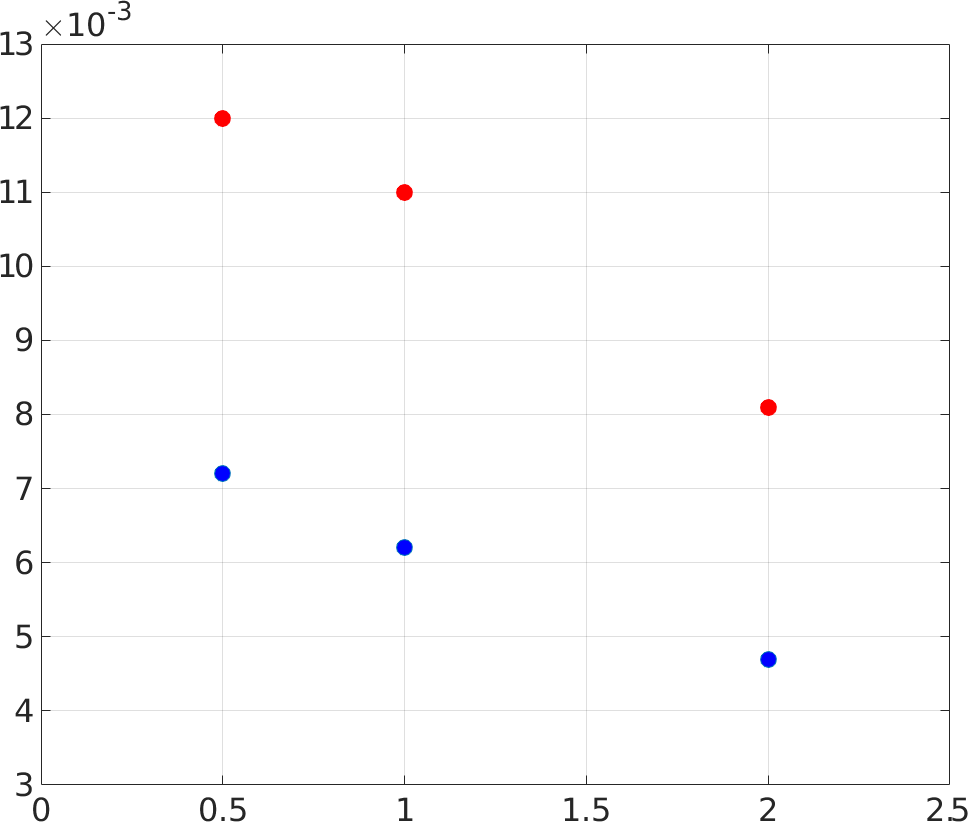}}
{${L_x/L_z}$}{1mm}{\begin{rotate}{90}${\dot \delta _\theta}$\end{rotate}}{3mm}
\end{minipage}
\hspace{0.6cm}
\begin{minipage}[c]{.43\linewidth}
\FigureXYLabel{\includegraphics[width = .9\linewidth]{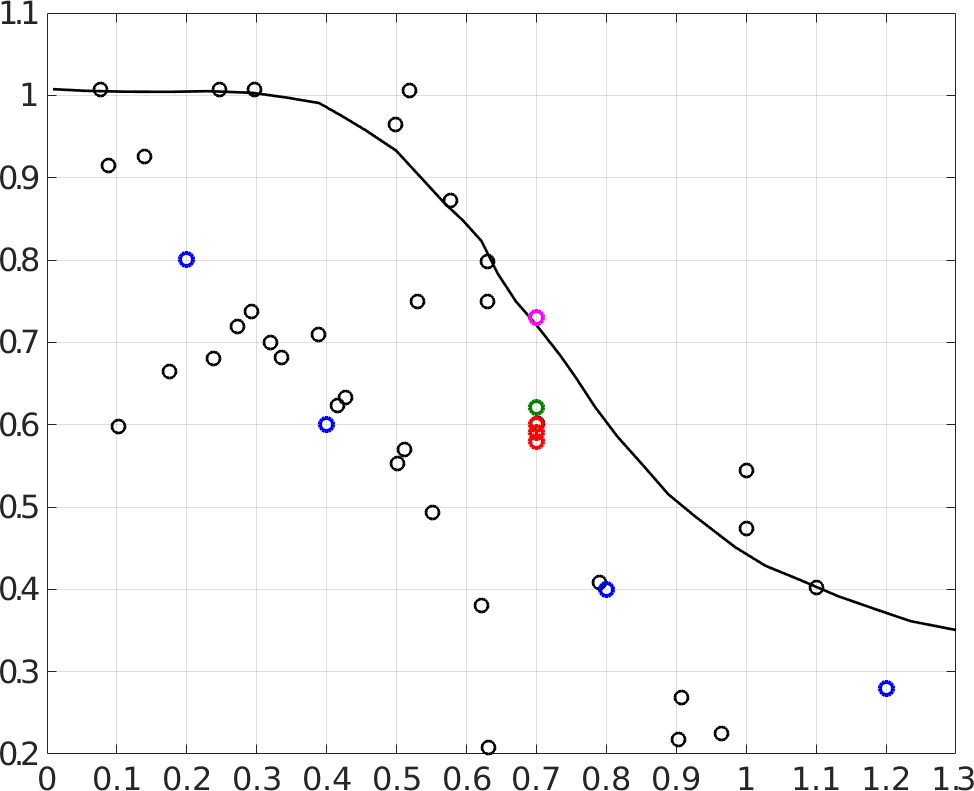}}
{${M_c}$}{1mm}{\begin{rotate}{90}${\dot \delta _{\theta,0.7}}/{\dot \delta_{\theta,0.3}}$\end{rotate}}{3mm}
\end{minipage}
\caption{ Left plot: Growth rates  vs. simulation box ratios for $M_c=0.3,~0.7$, red and 
  black points correspondingly; Right plot: Growth rate ratios from the different experimental
  and numerical studies of the flow in terms of convective Mach number.
Black circles, see \citet{Pantano2002,Foysi2009,Mitsuno2020} for
references and details. The blue circles correspond to the recent
results of Matsuno and Lele (2020) \cite{Mitsuno2020}. The red circles
are from the present simulations for different $L_x/L_z$ ratios. The green and magenta circles correspond to the cases A7 from Pantano and Sarkar, 2002 \cite{Pantano2002} and LES07 from Foysi and Sarkar, 2010 \cite{Foysi2009}, correspondingly. The black solid line corresponds to the Langley experimental curve.} 
\label{LC}
\end{figure}
The growth rates  vs. simulation box ratios ($L_x/L_z$) for $M_c=0.3$ and $0.7$ with red 
and black points, correspondingly are shown on left plot in Figure \ref{LC}. 
As one can see, the increase of box ratio leads to the decrease of  the growth rate for both Mach numbers.  
The growth rate ratios (${\dot \delta _{\theta,0.7}}/{\dot \delta_{\theta,0.3}}$) for 
different experimental and numerical studies are presented  on the right plot in Figure \ref{LC}. 
The blue filled circles correspond to the recent results for different Mach numbers by 
Matsuno and Lele (2020). The black filled circles represent different experimental and
numerical results publishes in the literature (details for each points can be found in the literature, 
see, for example the papers  by \citet{Pantano2002,Foysi2009,Mitsuno2020} and references therein).
As one can see the growth rate ratios of our simulations (red open
circles) are in good agreement with those available in 
the literature. Here it is worth to note that for calculation of the growth rate ratios, 
we used  the same simulation box ratios $L_x/L_z$ for the both Mach numbers. 
As one can see, in the case of quasi-incompressible case, at $M_c = 0.3$, 
the increase of box ratio leads to  the growth rate decrease that is clearly seen on the 
left plot in Figure \ref{LC}.
The mean profiles and  RMS of streamwise velocity for aspect ratios
corresponding to Cases 1-3 in self-similar period of flow
development are presented in Figure \ref{ss_R11}.  The turbulent
stress tensor is defined as $R_{ij}= \overline{\rho u^{\prime
\prime}_iu^{\prime \prime}_j}/\overline{\rho}$, with Favre
fluctuations \cite{Pantano2002,Foysi2009}. The collapse of Reynolds
stresses and mean velocities is observed as it is expected in the
self-similar period.  As one can see from the figure, the peak
values of the RMS of streamwise velocities decrease with increasing
of the aspect ratios of the simulation boxes.

\begin{figure}[H]
\centering
\centering
\begin{minipage}[c]{.43\linewidth}
\FigureXYLabel{\includegraphics[width = .9\linewidth]{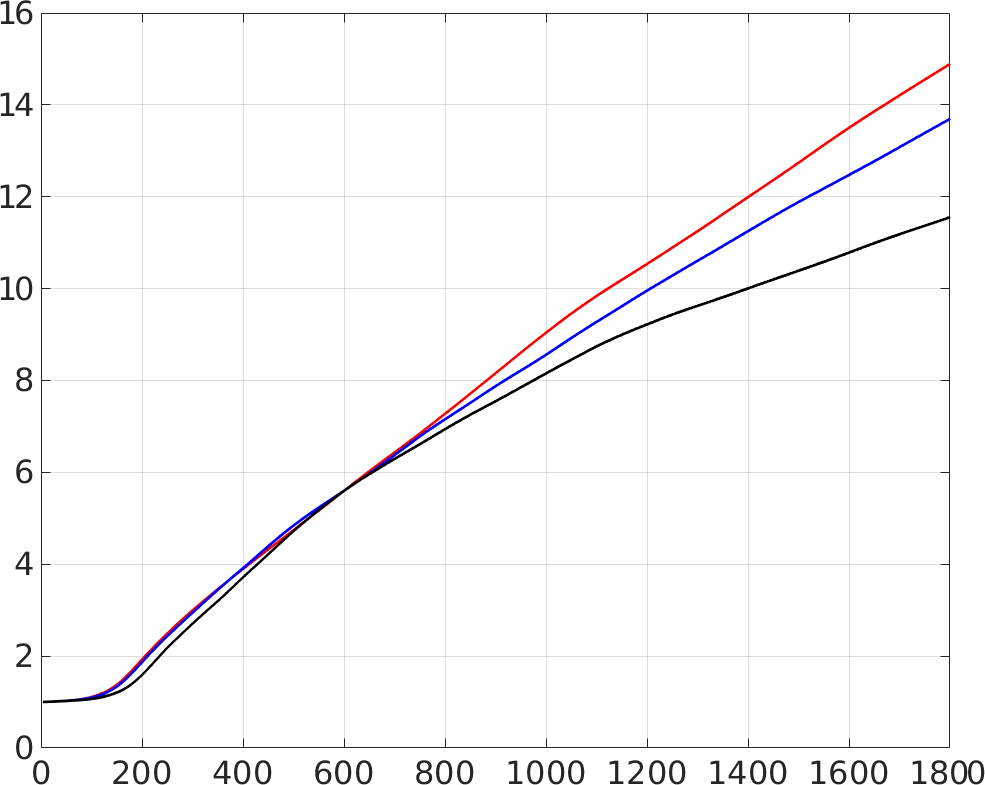}}
{${\tilde \tau}$}{1mm}{\begin{rotate}{90}${\delta _\theta}/{\delta _{\theta} (0)}$\end{rotate}}{3mm}
\end{minipage}
\caption{Evolution in time of the momentum thickness normalized on the initial value, 
  $\delta_{\theta}({\tilde \tau})/\delta_{\theta}(0)$, for different aspect ratios of
simulation boxes $L_x/L_z$: $0.5$ (red line), $1.0$ (blue line), and
$2.0$ (black line), from top to bottom.}
\label{thetas_main}
\end{figure}

The aim of the first three cases of simulations is to show the main
results of the work, to identify the origin of the acoustic wave
output.  The main difference between these cases is the different
streamwise-spanwise aspect ratios ($L_x/L_z$) of the simulation
boxes that define the characteristics of large scale
eddies/harmonics of the turbulence. For instance, in Case 2 the
mixing layer turbulence, in comparison with Case 1, contains a large
scale harmonic with the wavelengths $\lambda _x=100, \lambda_z=100$.
In Case 3, furthermore, the turbulence is even richer, additionally
containing the harmonic with $\lambda _x=100, \lambda_z=200$ that is
quite active in acoustic wave generation process. These different
spectral contents of the turbulence in Cases 1-3 slightly influence
the establishment of the self-similar state during the simulations,
which can be seen in Figure \ref{thetas_main}, showing  the momentum
thicknesses, $\delta{_\theta}({\tilde \tau})$, vs. the normalized
time ${\tilde \tau}=t \Delta U / {\delta{_\theta} (0)}$. In the
self-similar region (${\tilde \tau}> 1300$), the mixing layer grows
linearly with the growth rates depending on the aspect ratios of the
simulation boxes: $\dot {\delta_{\theta}}({\tilde \tau})  =
0.007,0.006,0.005$ for $L_x/L_z = 0.5,1,2$, respectively. It is
clear that the growth rate ${\delta_{\theta}}({\tilde \tau})$
decreases with increasing aspect ratio.

\begin{figure}[H]
\centering
\begin{minipage}[c]{.45\linewidth}
\FigureXYLabel{\includegraphics[width = .9\linewidth]{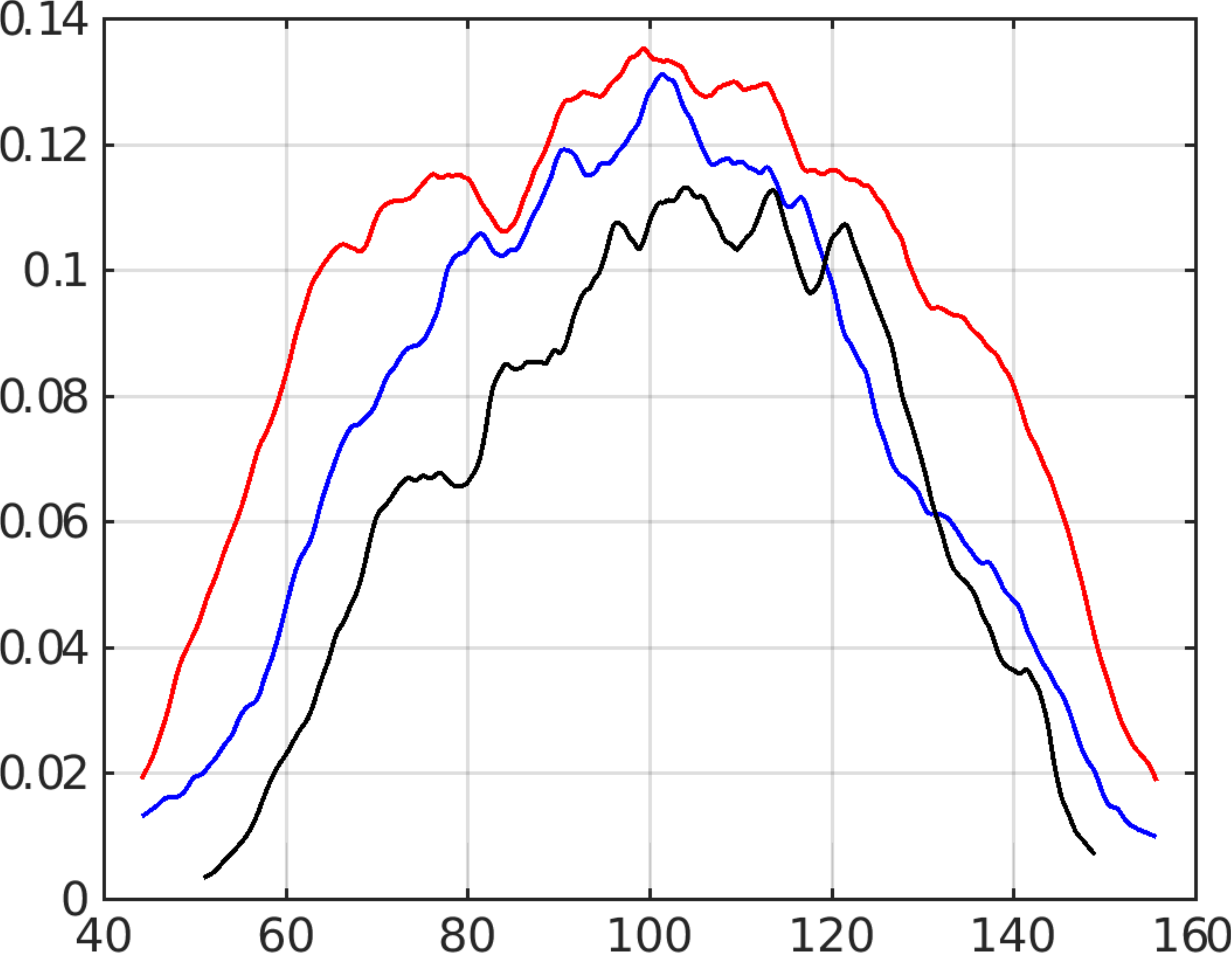}}
{$y$}{1mm}{\begin{rotate}{90}$\sqrt{R_{11}}/\Delta U$\end{rotate}}{2mm}
\end{minipage}
\hspace{0.5cm}
\begin{minipage}[c]{.45\linewidth}
\FigureXYLabel{\includegraphics[width = .9\linewidth]{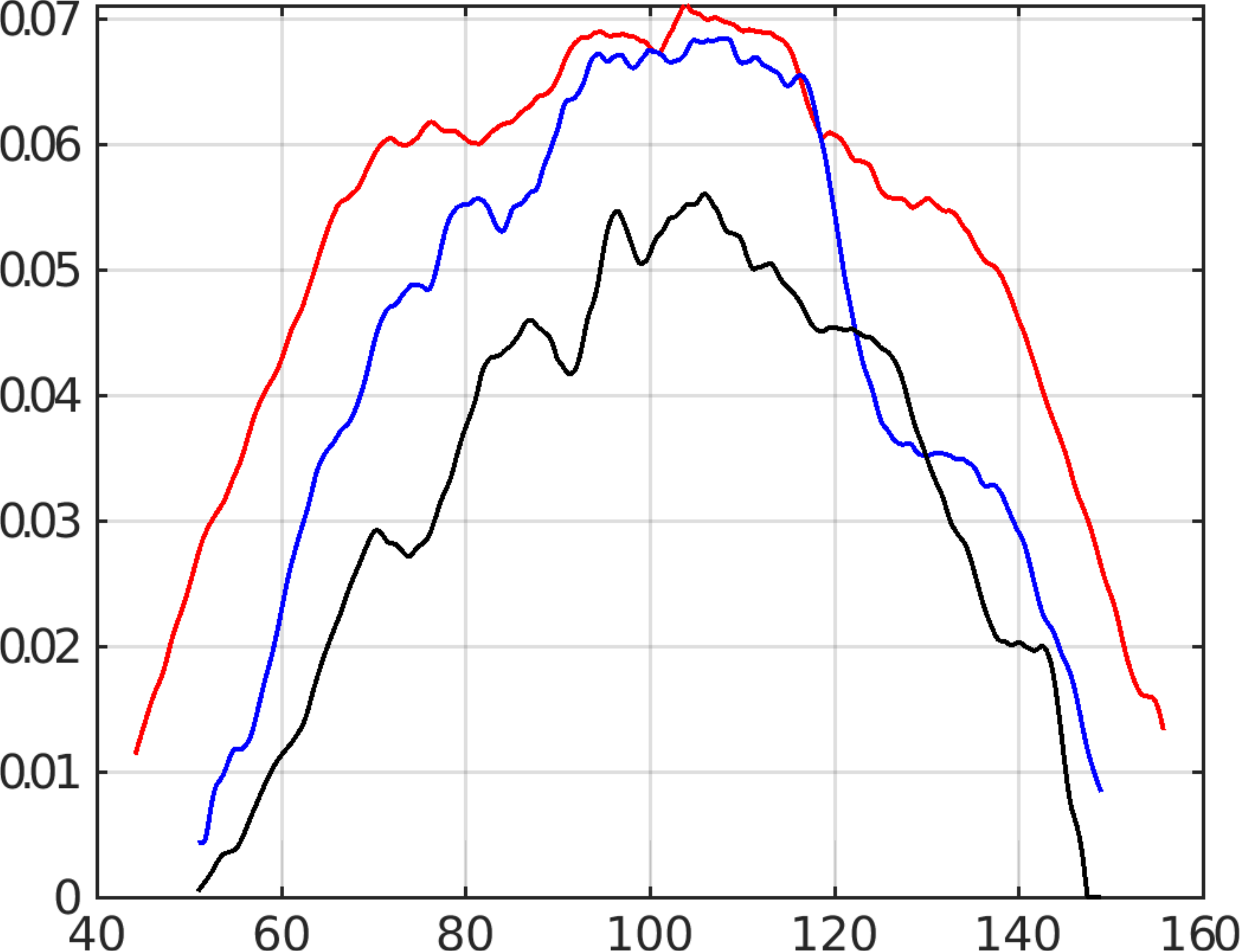}}
{$y$}{1mm}{\begin{rotate}{90}$\sqrt{R_{12}}/\Delta U$\end{rotate}}{2mm}
\end{minipage}
\caption{Profiles of the RMS-velocity and shear stress for different
aspect ratios of simulation boxes $L_x/L_z$: $0.5, 1, 2$, dashed red, fine dashed blue
and solid black lines correspondingly in the self-similar state at
${\tilde \tau} = 1600$.}
\label{stresses}
\end{figure}
\begin{figure}[H]
\centering
\begin{minipage}[c]{.43\linewidth}
\FigureXYLabel{\includegraphics[width = .9\linewidth]{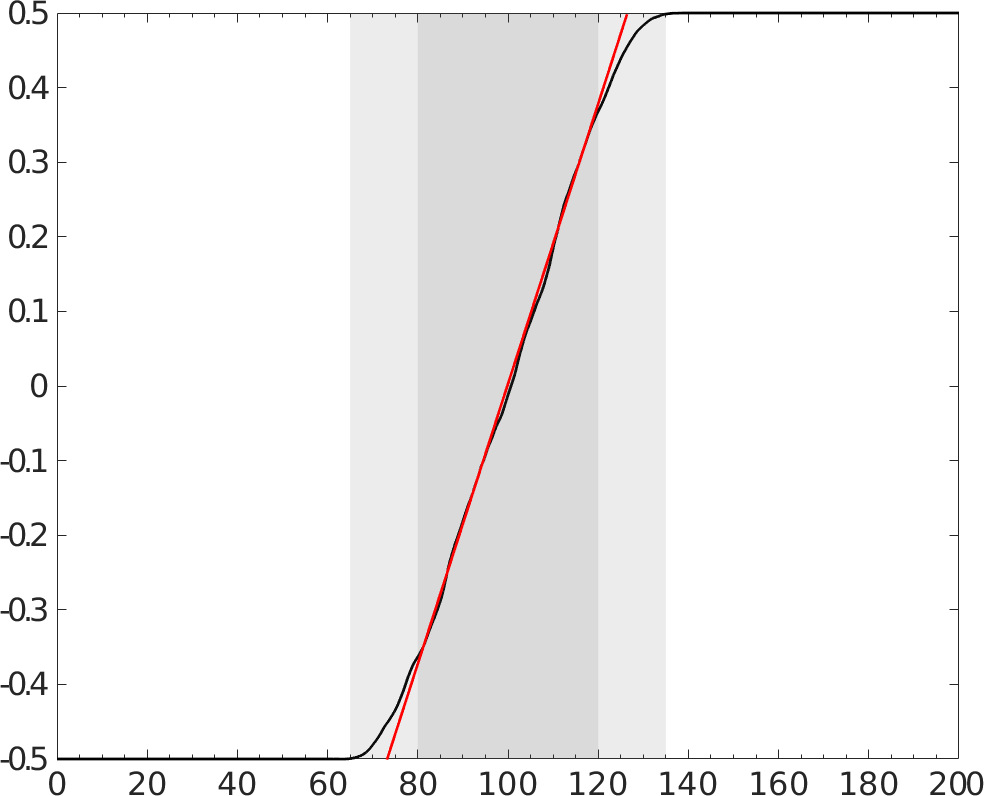}}
{$y$}{1mm}{\begin{rotate}{90}$\overline{U}$\end{rotate}}{3mm}
\end{minipage}
\hspace{0.5cm}
\begin{minipage}[c]{.43\linewidth}
\FigureXYLabel{\includegraphics[width = .9\linewidth]{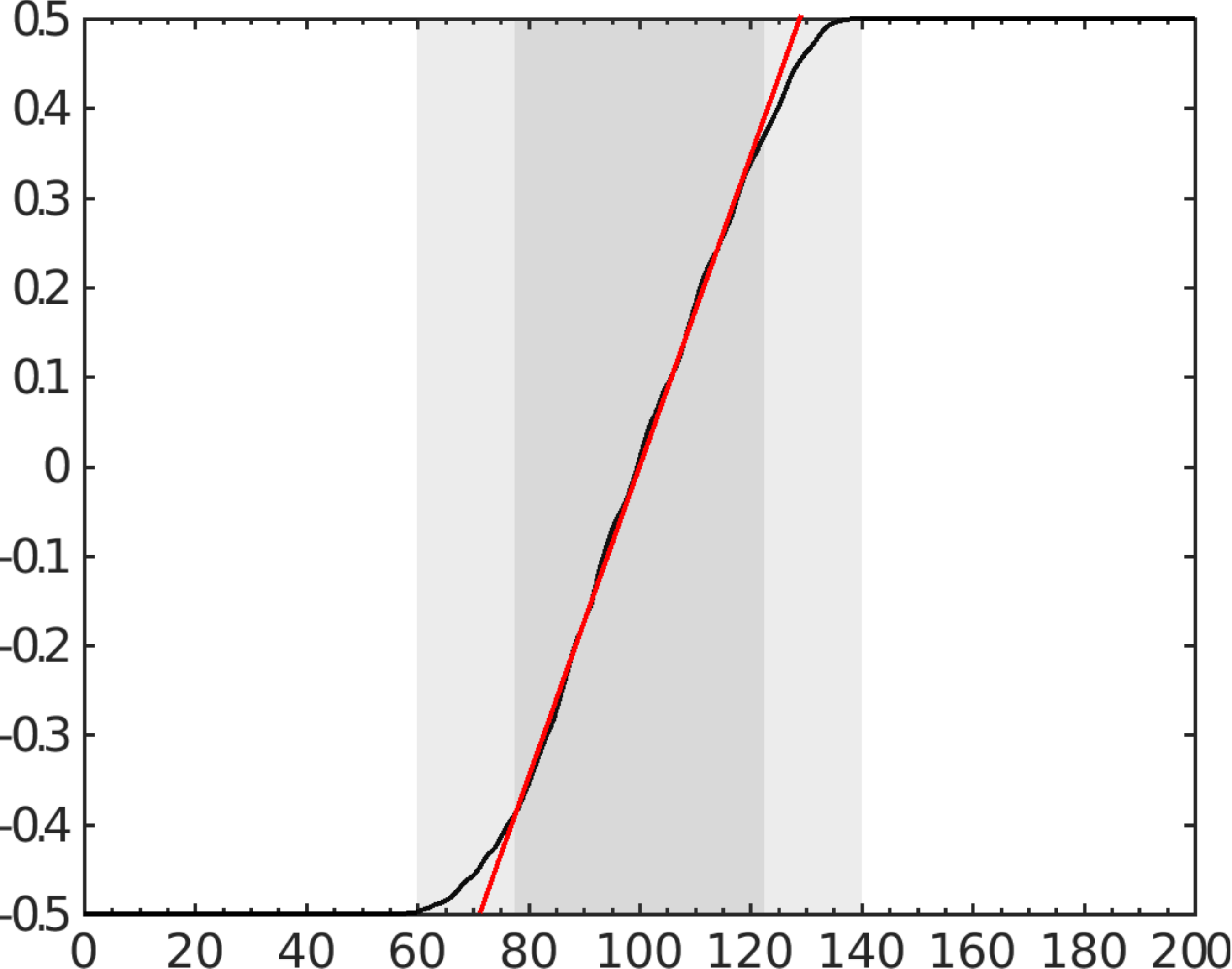}}
{$y$}{1mm}{\begin{rotate}{90}$\overline{U}$\end{rotate}}{3mm}
\end{minipage}
\caption{Mean velocity profiles (the black line) in the self-similar
period of the flow evolution for $M_c =0.3$ (left plot) and $M_c =0.7$ (right plot) 
are shown at ${\tilde \tau} \approx 930$ and ${\tilde \tau} \approx 1300$, correspondingly. 
The red line shows the maximum gradient of the mean velocity. The grey area represents 
the turbulent region of the flow. The dark grey area corresponds to the central/body 
region of the flow with linear mean velocity profile which matches to the red
line. So, the red line inclination  defines the shear rate, $A(\tilde \tau)$, in the 
body region -- the region of the acoustic wave mode generation. In the marginal, 
right and left white areas the  near  acoustic wave fields (generated in the central 
area and then emitted from the shear layer) are formed.}
\label{Um_ss}
\end{figure}

Figure \ref{stresses} shows the RMS of the streamwise velocity and
shear stress for different aspect ratios of the boxes. These
quantities are presented in a fixed moment of time (i.e., without
 averaging in time), therefore, they have not a very regular look.
The figure shows significant influence of the box ratio on the
turbulence intensity -- it decreases with increase of $L_x/L_z$.
Figure \ref{Um_ss} shows the geometrical characteristics of the
shear layer in the self-similar region, at ${\tilde \tau} \approx
1300$, for Case 1 ($M_c=0.7$, right plot) and at 
${\tilde \tau} \approx 930$, for Case 4  ($M_c=0.3$, left plot). 
The central region (with constant shear rate)
is more than 50\% of the mixing layer total width. In addition, in this
region, the shear rate substantially exceeds the rate in the rest
of the layer. Therefore, it is natural that this region is dominant
in the generation process of acoustic waves and, ultimately, in the
formation of the near wave field.

According to the view presented in Sect II, the basic ``elements''
of the linear wave generation mechanism in constant shear flows are
the Kelvin modes -- the harmonics of the perturbations. Therefore,
below we analyse the acoustic wave generation in terms of Kelvin
modes in the central part of the shear layer.

To grasp the essence of our view on the wave generation process and
apply it to the  compressible turbulent time-developing shear layer,
one should understand the difference and define quantitative
relation between $M_c$ and ${\mathcal {M}}$. The former is
determined only through general flow parameters, whereas the latter
depends on the parameters of the central part of the shear layer and
its perturbation harmonics as ${\mathcal {M}}=A/(k_x
c_s)=A{\lambda}_x/2\pi c_s$, where ${\lambda}_x$ is the streamwise
wavelength of the harmonic and $A$ is the shear rate. One can
determine $A$ from the inclination of the red lines in Figure
\ref{Um_ss}: $A=\Delta U/\Delta y$. Similarly, two different
nondimensional times connected to the wave harmonics, $\tau=c_sk_xt$
and to the shear layer, ${\tilde \tau}=t \Delta U / {\delta_\theta
(0)}$ have been introduced. Taking Equation \eqref{Mc} into account, one gets
the following relation between the Mach numbers:
\begin{equation}
\label{M} 
{\mathcal {M}} = \frac{{\Delta U} \lambda_x}{2\pi c_s \Delta y}=
\frac{M_c \lambda_x}{\pi \Delta y}.
\end{equation}
From this equation we can get the following values of the mode-dependent 
Mach number ${\cal M}$ for $M_c = 0.3, ~0.7$ in the self-similar period 
of flow development, ${\tilde \tau} \approx 930, ~1300$,
correspondingly:
\begin{equation}
\label{M1} 
{\mathcal {M}^{0.7}} = \frac{0.7 \lambda_x}{\pi 60} ~~~ \text{and} ~~~ 
{\mathcal {M}^{0.3}} = \frac{0.3 \lambda_x}{\pi 50}.
\end{equation}

We analyze the acoustic wave generation in the self-similar period where the growth rate of the momentum
thickness remains constant, i.e., at ${\tilde \tau}>1300$
(see Figure \ref{thetas_main}). Below we present and analyze the
wave generation and dynamics specifically at ${\tilde \tau} \approx
1600$ for $M_c=0.7$. At first, we calculate the mode-dependent Mach number for
the first/largest streamwise harmonics with the wavelength equal to
the box streamwise size, $\lambda_{x1}=L_x=100$. From Equation
\eqref{M1}, for $M_c=0.7$, the
mode-dependent Mach number for the first/largest streamwise harmonic
can be calculated: ${\mathcal M}^{0.7}(\lambda_{x1})\approx 0.372$.
From the equation it follows that the
mode-dependent Mach number decreases proportionally with
$\lambda_{x}$. Therefore, for the second harmonic, $\lambda_{x2}=50$, 
it is twice less, ${\mathcal {M}}^{0.7}(\lambda_{x2}) \approx 0.186$, etc. 
From the right plot in Figure \ref{fig:eta} one can see that the wave generation is significant at
$\mathcal {M}^{0.7}\approx 0.372$, but it is negligible  at $\mathcal {M}^{0.7}
\approx 0.186$: $\eta(\mathcal {M}^{0.7} \approx 0.372)\approx 0.235$,
$\eta(\mathcal {M}^{0.7} \approx 0.186)\approx 0.0069$, i.e., $\eta(\mathcal
{M}^{0.7} \approx 0.372)/\eta(\mathcal {M}^{0.7} \approx 0.186) \approx 34$. Thus,
in the analyzed shear layer with $M_c=0.7$ and considered simulation
boxes with streamwise length $L_x=100$, there occurs a substantial
generation only for acoustic wave harmonics with the largest streamwise
wavelength $\lambda_{x1}=100$. 

In similar way, the wave generation for $M_c=0.3$ can be analyzed.  
From Equation (\ref{M1}) we have: ${\mathcal M}^{0.3}(\lambda_{x1})\approx 0.191$.
Consequently, the ratio between high and low Mach numbers for $\lambda_{x1}=100$ is 
$\eta(\mathcal {M}^{0.7} \approx 0.372)/\eta(\mathcal {M}^{0.3} \approx 0.191) \approx 18$, 
i.e. the flow is, in fact, quasi-incompressible at $M_c = 0.3$.

\begin{figure}[H]
\centering
\includegraphics[width = .28\linewidth]{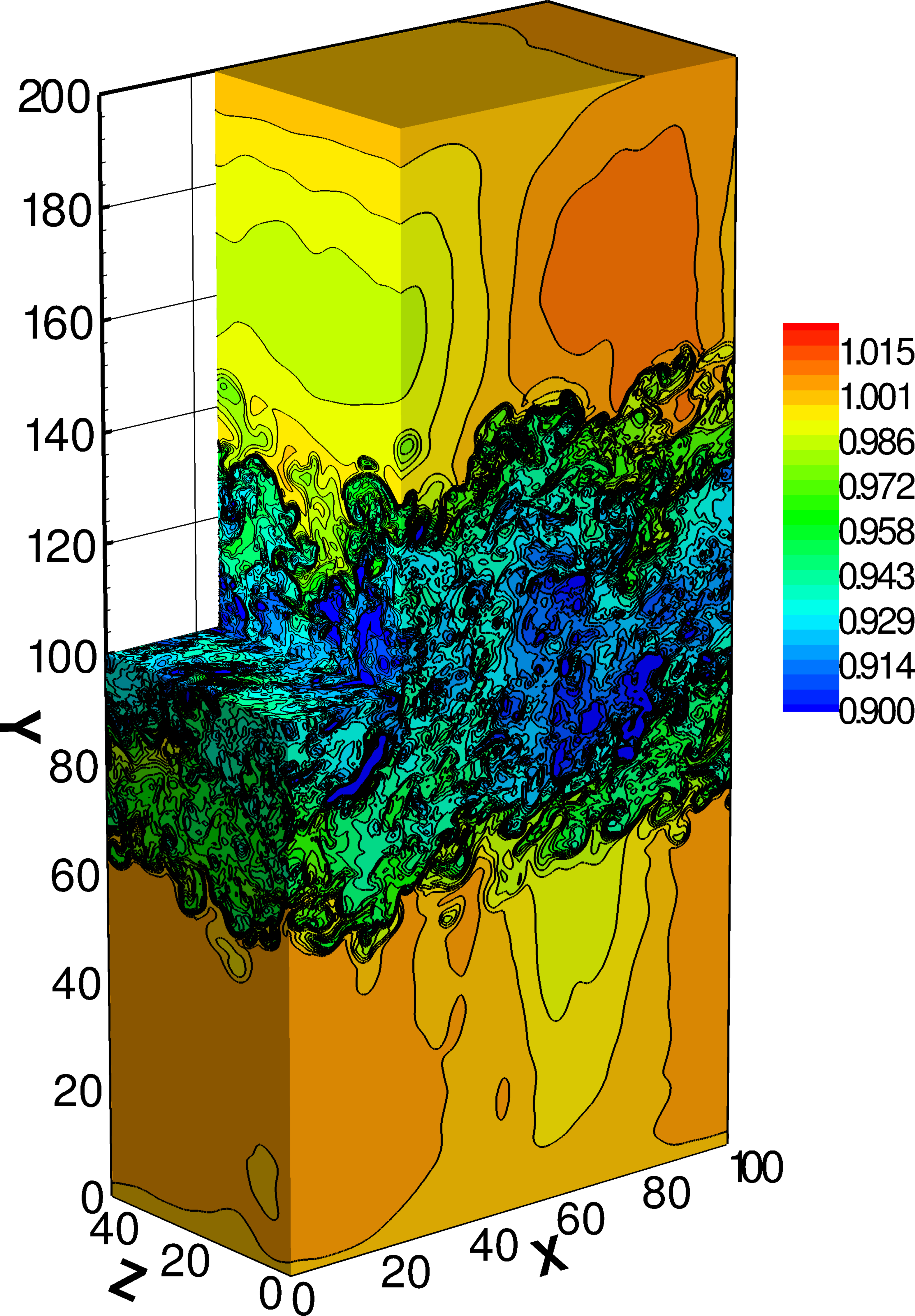}
\hspace{0.3cm}
\includegraphics[width = .42\linewidth]{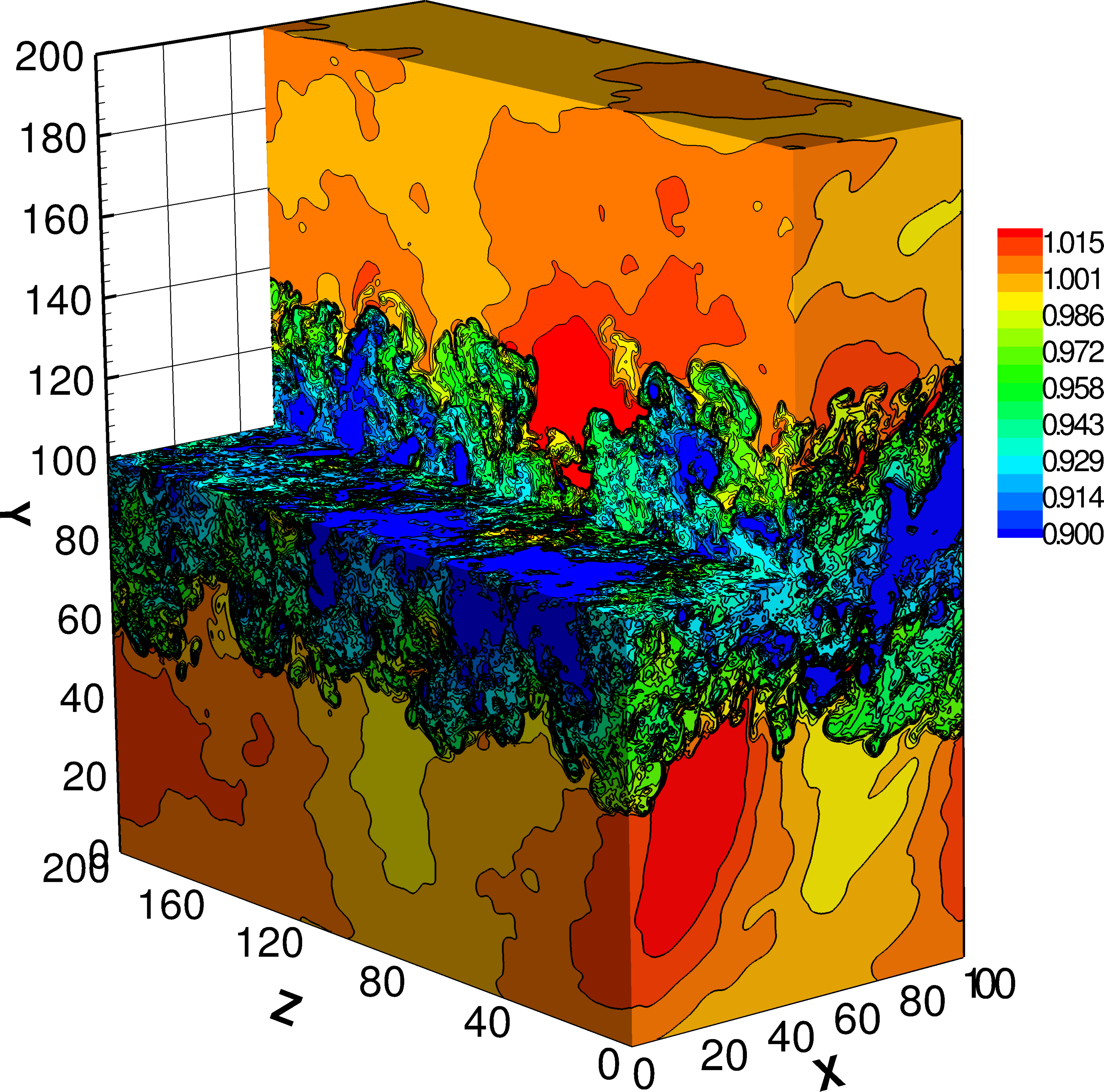}
\caption{3D surfaces of the density field in the self-similar stage
of the flow is shown in the turbulent mixing layer area and near far
acoustic wave field areas. The left plot with the well pronounced 2D
near  field represents Case 1 ($L_x/L_z=2$) and the right one
with the well pronounced 3D near  field shows Case 3
($L_x/L_z=0.5$). The near  field configurations (iso-surfaces)
for different aspect ratios are separately presented on Figures
\ref{far_fields1_2} and \ref{far_fields05}.}
\label{3D_fig3}
\end{figure}
\begin{figure}[H]
\centering
\begin{minipage}[c]{.4\linewidth}
\FigureXYLabel{\includegraphics[width = 1.\linewidth]{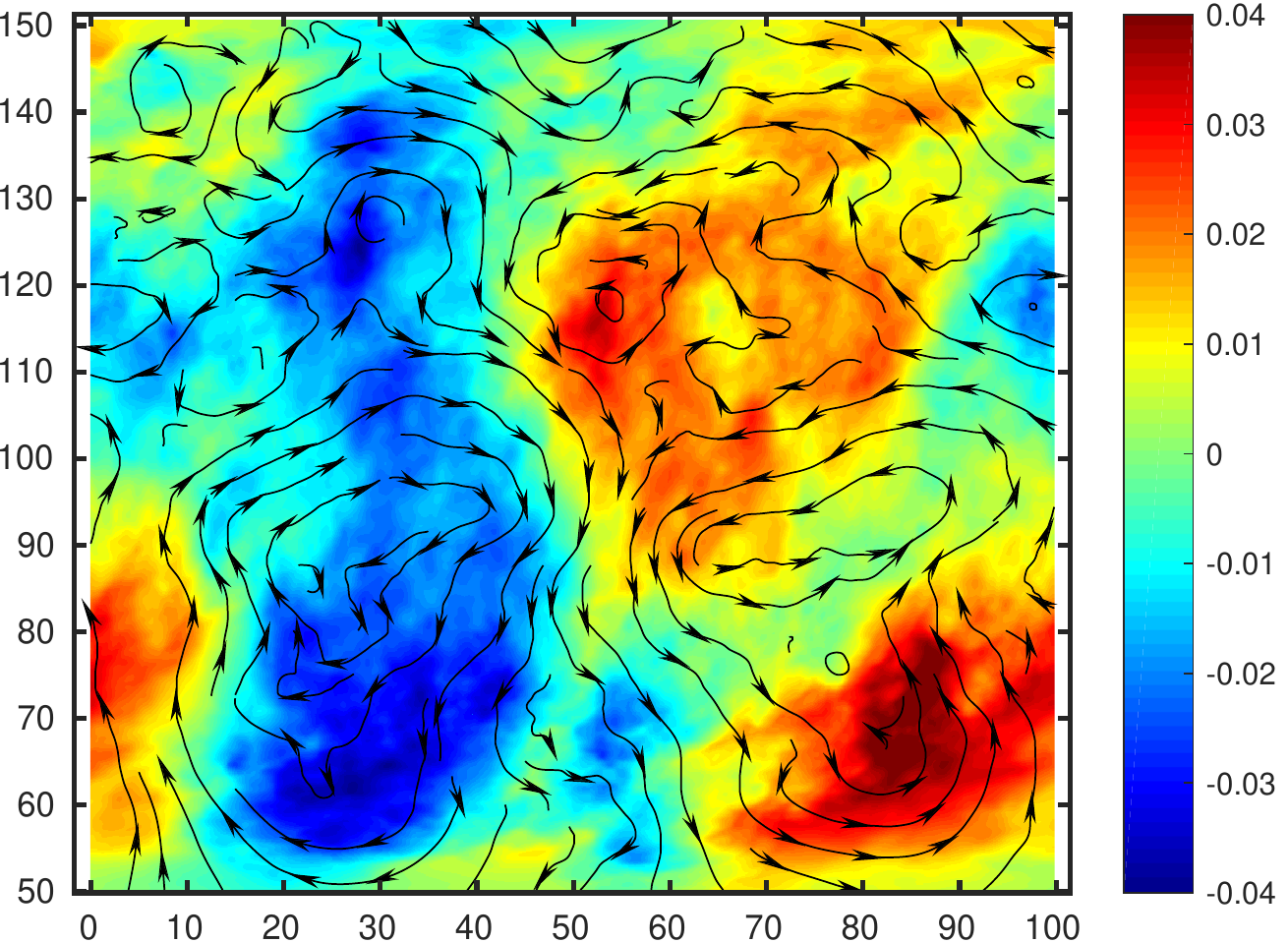}}
{$x$}{-1mm}{\begin{rotate}{90}$y$\end{rotate}}{1mm}
\end{minipage}
\caption{$xy$-slice of spanwise averaged density and velocity
streamtraces in the shear layer region ($50 < y < 150$) for Case 2
($L_x/L_z=1$).}
\label{core_region}
\end{figure}
For ${\tilde \tau} \approx 1600$, which corresponds to an
instant within the self-similar development of the flow -- the increase of the shear
layer thickness occurs with a constant rate. The turbulence
is expanding and ``fills'' the layer. Therefore, the linearly generated
acoustic waves immediately appear outside the layer. Figure
\ref{3D_fig3}, that represents the simulation Case 1 with
$L_x/L_z=2$, shows that the shear layer area generally consists of 3D
vortical turbulence, while the emitted acoustic field is mostly 2D.
Obviously, the 3D turbulent field of the shear layer contains 2D
components responsible for the generation of 2D acoustic waves.
Spanwise averaging of the density and velocity perturbation fields,
presented in Figure \ref{core_region}, reveals such 2D components in the
mixing layer central region ($50 < y < 150$). As one can see, mainly
the vortex component with maximum wavelength ($\lambda_{x}=100$) is
pronounced, although lower harmonics ($\lambda_{x}=50,25$) are also
traced that weakly violate the regular patterns of the largest
harmonic. From this ensemble of 2D vortex harmonics (for the
considered convective Mach number, $Mc=0.7$) the dominant
contribution to the generation of acoustic waves (as it is
calculated above) is given by the largest one.

\begin{figure}[H]
\centering
\begin{minipage}[c]{.4\linewidth}
\includegraphics[width = .9\linewidth]{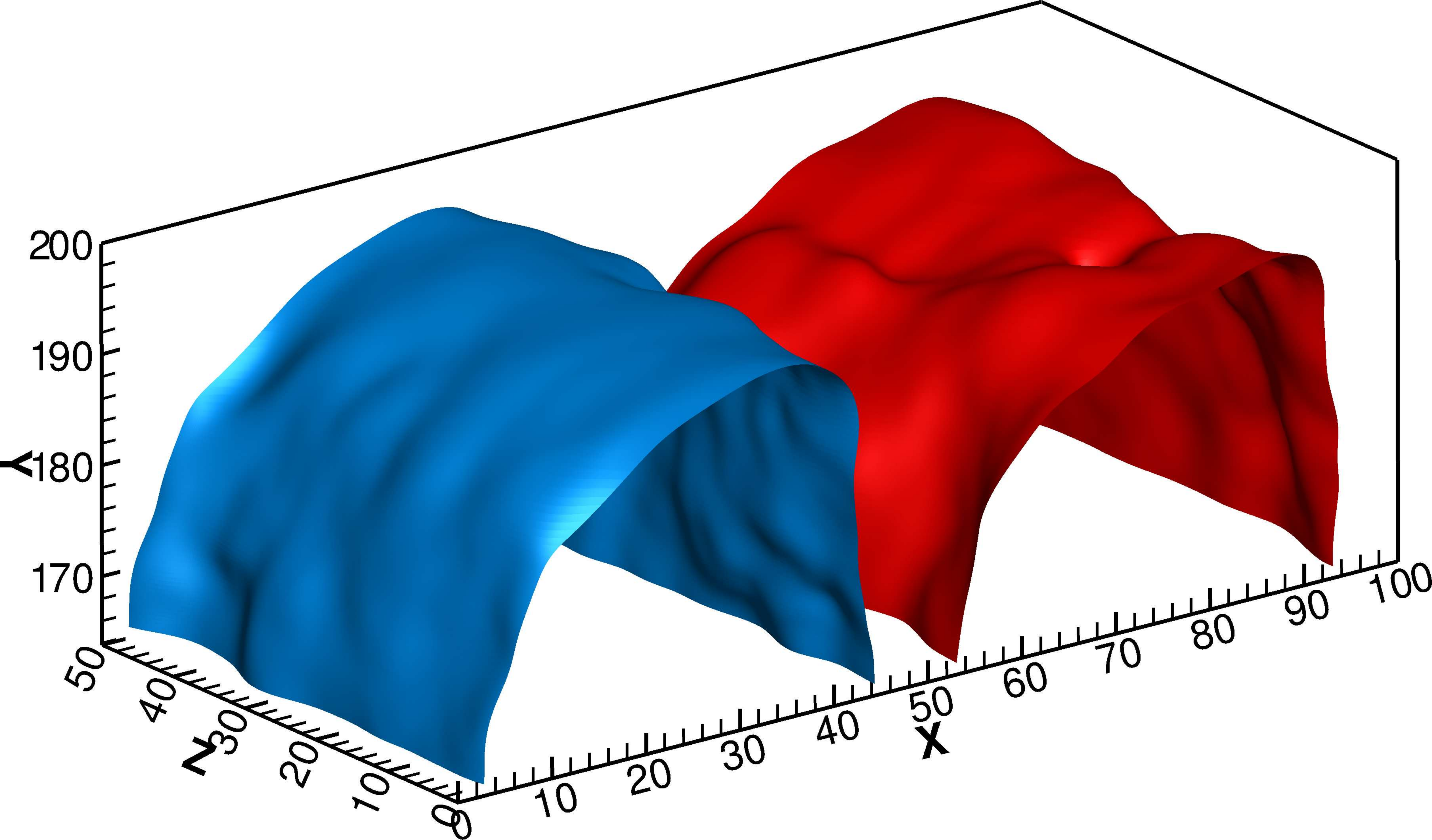}
\end{minipage}
\hspace{0.3cm}
\begin{minipage}[c]{.5\linewidth}
\includegraphics[width = .9\linewidth]{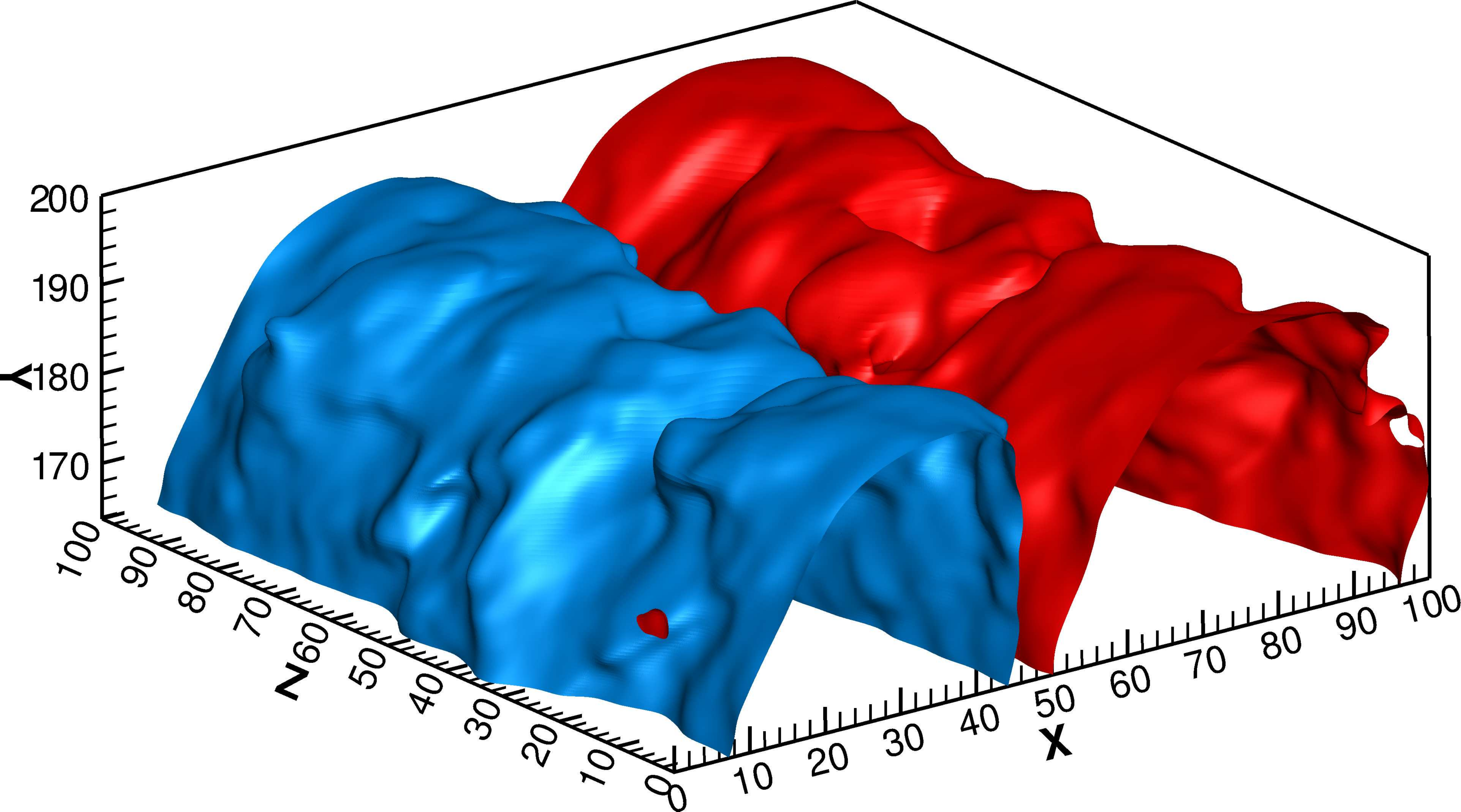}
\end{minipage}
\caption{Iso-surfaces of the near  density fields with $\rho = 0.996$ (blue) and
$\rho = 1.004$ (red). The left and right plots correspond to Case 1 and Case 2 correspondingly.}
\label{far_fields1_2}
\end{figure}

According to Equation \eqref{eq:M_max} the mode-dependent Mach
number for 3D harmonics, $\mathcal {M}^*$, decreases with increase of $\gamma$.
Consequently, the wave generation also decreases with increase of $\gamma$.
In Case 1, for the harmonic with the largest streamwise wavelength, the minimal
value of spanwise normalized wavenumber is $\gamma=\lambda_{x}/\lambda_{z}=L_x/L_z=2$.
The calculated efficacy of the wave generation for this 3D harmonic is
$\eta(\lambda_{x}=100,\lambda_{z}=50)=3.2\cdot10^{-7}$ which is
negligible compared to $\eta(\lambda_{x}=100,\lambda_{z}=0)=0.32$.
This explains the pronounced 2D nature of the acoustic wave density
near  field outside the shear layer. The 2D
structure of this field is even more strongly pronounced in Figure
\ref{far_fields1_2} where the near  field for two
values of density, corresponding phases of compression ($\rho =
1.004$, red color) and rarefaction ($\rho = 0.996$, blue color) is presented. The
left and right plots represent Case 1 and Case 2 correspondingly.
The 2D structure of the near  wave field is less established in Case 2
(there is an additional admixture of 3D wave harmonics), where the minimal
value of spanwise normalized wavenumber is $\gamma=\lambda_{x}/\lambda_{z}=L_x/L_z=1$.

\begin{figure}[H]
\centering
\begin{minipage}[c]{.6\linewidth}
\includegraphics[width = .9\linewidth]{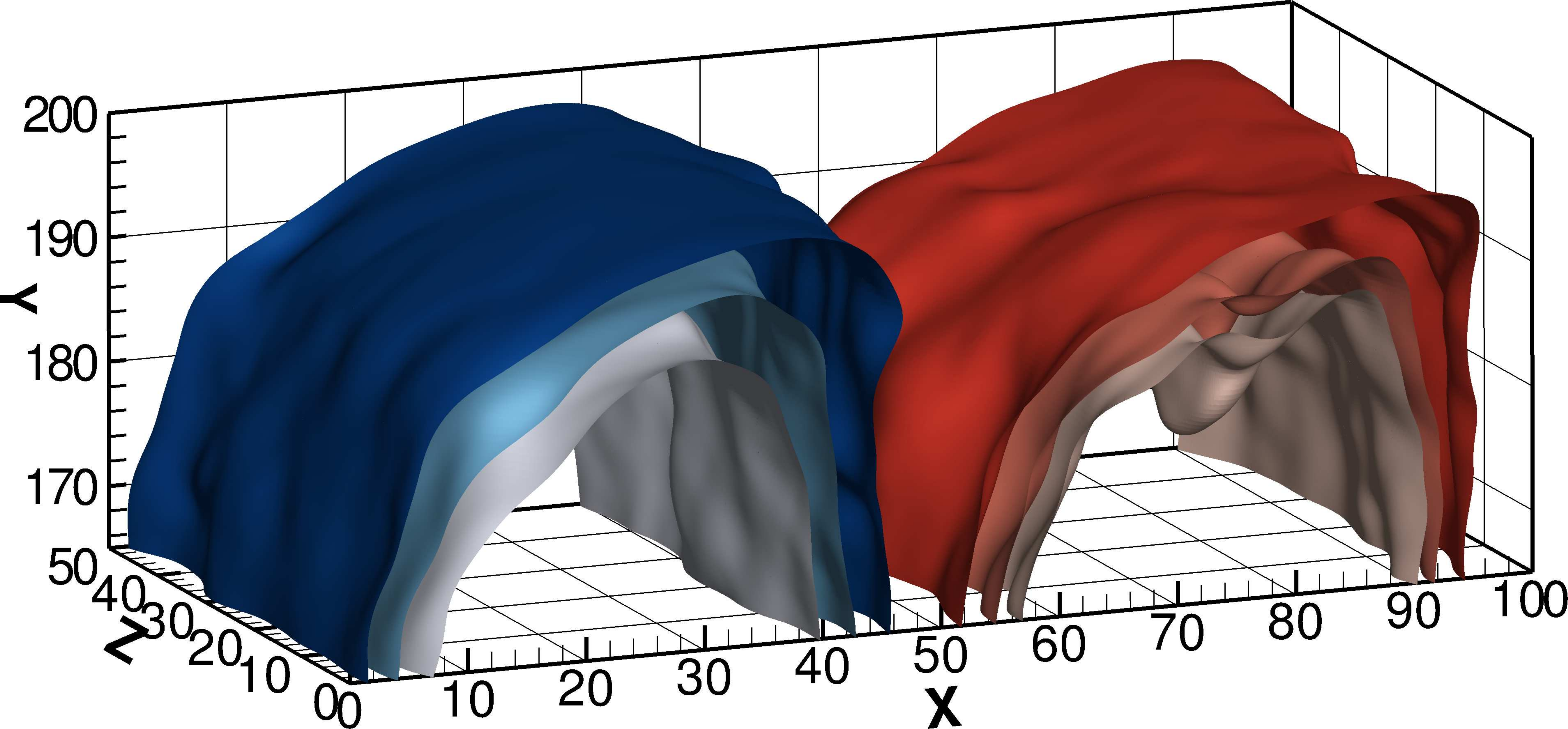}
\end{minipage}
\caption{Iso-surfaces of the near  density fields with $\rho = 0.992,0.994,0.996$
(blue color group) and $\rho = 1.002,1.004,1.006$ (red color group)
at $L_x/L_z = 2$.}
\label{far_fields_multi}
\end{figure}

For clarity we present one more Figure \ref{far_fields_multi} for
Case 1. It shows three iso-surfaces of the near  density
field. The left group of iso-surfaces of the blue color corresponds to three surfaces of
the density in the  rarefaction phase ($\rho = 0.992, 0.994, 0.996$)
and the right group (red color) corresponds to three surfaces of the
density in the compression phase ($\rho = 1.002, 1.004, 1.006$).
Some visible irregularity of the right inner surfaces are traces of
weak subsequent harmonics.

\begin{figure}[H]
\centering
\begin{minipage}[c]{.7\linewidth}
\includegraphics[width = .9\linewidth]{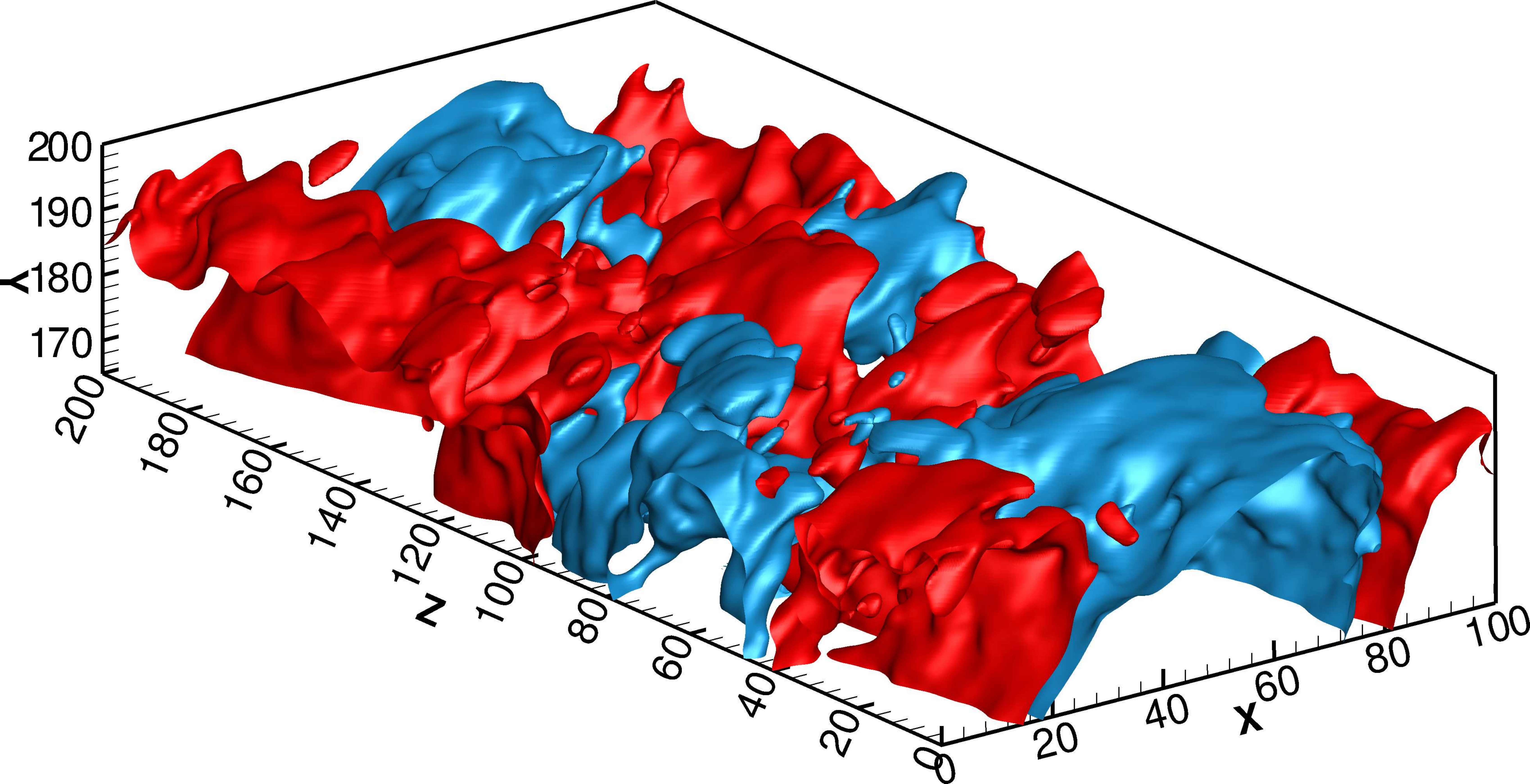}
\end{minipage}
\caption{Iso-surfaces of the near  density fields with $\rho =
0.996$ (blue) and $\rho = 1.004$ (red) at $L_x/L_z = 0.5$.}
\label{far_fields05}
\end{figure}

The picture of the density field is fundamentally different in
Case 3, where the lower minimal value of the spanwise normalized wavenumber for 
the harmonic with the largest streamwise wavelength is
 $\gamma=\lambda_{x}/\lambda_{z}=L_x/L_z=100/200=0.5$.
 The calculated efficacy of the wave generation for this largest 3D harmonic is
$\eta(\lambda_{x}=100,\lambda_{z}=200)=0.22$ that is comparable to
the efficacy of the related 2D harmonics $\eta(\lambda_{x}=100,\lambda_{z}=0)=0.32$
(compare also the amplitudes of the black and red lines of the right column of Figure \ref{lin_case}).
Therefore the near  density field represents a mix of these 2D and 3D harmonics.
This complex picture is containing a significant 3D component in the near 
field and is shown in Figure \ref{far_fields05}, where (as in Figure
\ref{far_fields1_2}) the near  field of the density corresponding to phases of compression
($\rho = 1.004$, red color) and rarefaction ($\rho = 0.996$, blue color) is presented.

Summarizing the results of DNS we can say that the near  acoustic
field exactly matches the characteristics of linearly generated
acoustic waves in the central/body part of the shear layer due to
the vortex--wave mode coupling induced by the shear flow non-normality
as described in section \ref{sec:model_equation}. The linear generation
is substantial at ${\mathcal {M}}>0.4$. This condition is satisfied
by the harmonics of the considered shear layer (with $M_c=0.7$ and
$L_x=100$) with the largest streamwise wavelength, $\lambda_x=100$.
The dominance of the largest streamwise harmonic is clearly seen
from Figures \ref{far_fields1_2} and \ref{far_fields_multi}. These
figures correspond to Cases 1 and 2, for which the minimal values are
$\gamma=2$ and $1$, respectively. For these cases the efficacy of 3D
wave generation is small, $\eta \ll 1$ (see also Figure \ref{lin_case}),
therefore, the generated waves should be mostly 2D, which is clearly
demonstrated by these figures. The very regular and smooth picture
of the near  wave field, shown in Figures \ref{far_fields1_2} and
\ref{far_fields_multi}, is easily explained from the abrupt and very
regular character of the emergence of wave harmonics, according to
which at the moment of the emergence, the shearwise velocity and
density of the wave harmonic are zero, while the values of the
streamwise and spanwise velocities are maximal (see item \emph{(iv)}
in section \ref{sec:model_equation}). Due to this regularity, the
wave harmonics generated at all subsequent moments are
phase-correlated and interfere constructively. Therefore, ultimately,
they give rise to a very regular and smooth near  wave field
presented in Figures \ref{far_fields1_2} and \ref{far_fields_multi}.
This time coherence of the generated acoustic waves
is subsequently permanent in nature.

\section {Summary and discussions}
\label{sec:conclusion3D}
The guideline of our efforts is the breakthrough within the hydrodynamic
stability community in the comprehension of the shear flow non-normality
and mathematical description of the non-normality induced phenomena.
This trend formed in the hydrodynamic stability community in 1990s
has been successfully adopted in atmospheric and astrophysical flow
communities. A similar process is taking place in the fusion plasma
community. Overall, the non-normality induced phenomena are
inevitable in all -- spectrally stable and unstable -- shear flows
as they essentially change the finite time period of the
perturbation dynamics. The non-normality brings forth a new type of linear
phenomenon -- vortex--wave mode coupling that ensures the emergence
of quite strong wave harmonics with very specific and regular
initial characteristics. The key factor of the non-normality of the shear flow's linear
operators is demonstrated in our work on time-dependent shear layers.
The main DNS result is the identification of the origin of the acoustic wave output,
specifically, the demonstration of the dominance of the linear generation process 
of acoustic waves from the
shear layer central part induced by the flow non-normality, on the acoustic wave near  field emitted from the shear layer. This certainly justifies the
importance of the shear flow linear operators non-normality and its
consequence - non-modal dynamics of the perturbations - in the formation of
acoustic wave output.

DNS and analysis performed for a specific shear flow give us an
idea of the particular role of the wave-packets in the emission of
acoustic waves from natural and engineering shear flows. One can
consider a shear flow where not only the largest streamwise vortex
harmonics are acoustically active, but also the subsequent shorter
ones. If these vortex harmonics are coherent in space they will form
a wave-packet. In this case, the spatial coherence of the
wave-packet would be added to the temporal coherence of the acoustic
wave generation mechanism described in section
\ref{sec:model_equation}. This would further amplify the generated
acoustic waves while maintaining their regular/smooth pattern and
eventually result in the formation of a regular/smooth near 
acoustic field which, in turn, evidently, transforms into the Mach
waves. At a first glance it all looks like wave-packet generated
acoustic waves. However, in fact, the waves are generated by the
vortex harmonics due to the mechanism induced by the shear flow
non-normality -- the wave-packet only collects the generated
acoustic wave harmonics by constructive interference of the
spatially and temporally coherent vortex harmonics of the
wave-packet.

The configuration of the considered mixing layer is quite simple and
permits to model the central part of the flow as a constant shear
flow. In case any modal (exponentially growing) instability is
absent -- the only linear mechanism of acoustic wave generation is
the vortex--wave coupling induced by the flow non-normality.
However, an exponential growth of acoustic waves occurs in other
kinds of engineering nonuniform flows with more complex velocity
profiles, for instance in jet flows (see, e.g. \cite{Jordan2013} and
references therein). Because of this, we consider it appropriate to
present the influence on the modal instability of the non-normality
of nonuniform flows when it occurs. \textit{In general, the
non-normality essentially changes the finite time period of any
type of dynamical processes in such flows.} This equally applies to
the linear transient phenomena in spectrally stable flows (e.g.,
transient growth of vortex mode perturbations, vortex--wave mode
coupling) and to spectrally unstable flows, where exponentially
growing ``modal solutions'' exist. From this the necessity follows
to take the non-modal effects on the exponentially growing mode
dynamics into account and not rely solely on its modal growth in
order to properly describe and understand their real dynamics. Let
us mention again, that this view about the importance of non-modal
physics in the dynamics of exponentially growing modes is well
comprehended by e.g. the astrophysical disk flow community
\cite{Squire_Bhattacharjee14,Gogichaishvili18}, too. Consequently,
the modal approach of exponentially unstable modes of wave-packets
in jet flows \citep{Jordan2013} can be misleading without involving
the modifications of the spectrally unstable mode dynamics induced
by the flow non-normality. The most justified way for nonuniform
flows with complex velocity profiles is to adapt the generalized
formulation of the non-modal approach
\cite{Farrell1996,Farrell1996a,Trefethen2005a,Schmid2007}, that
extends the modal stability theory to comprehensively account for
all transient processes in different kind of shear flows, including
the interaction between various modes and the mean flow, regardless
of whether the modes are spectrally stable or not. In any case,
further incorporation of phenomena induced by the shear flow
non-normality in the aerodynamic acoustic wave generation process
should be done in order to enhance the development of a refined
theoretical framework to guide noise-control efforts.

Finally, we present an additional weighty argument justifying the
importance of the linear vortex--wave coupling induced by the shear
flow non-normality. The specificity of the linear acoustic wave
generation mechanism substantiates the observations of two different
sound generating mechanisms, i.e., omnidirectional and highly
directional sound emission from fine-scale turbulence and large
turbulence structures, respectively \citep{Tam2008}. This is related
to nonlinear and linear mechanisms. The direct influence of
nonlinear (quadrupole) sources in the generation of acoustic waves
in the considered case is negligible -- to a greater degree acoustic
waves are generated in the central/body part of the shear layer due
to the vortex--wave mode linear coupling induced by the flow non-normality.
It should be mentioned, however, that an indirect influence of the
nonlinearity on the wave generation is significant -- nonlinear
phenomena take an active part in the formation of the turbulence
spectrum, including the large scale coherent structures which are
directly responsible for the generation of acoustic waves due to the
above described linear mechanism. The main topological feature in spectral space of the linear mechanism of acoustic wave harmonics generation by
vortex mode ones in constant shear flow is the following: the
generation occurs when vortex mode harmonics cross the line of $k_y
= 0$ in spectral space. Consequently, only up-shear tilted
($k_y/k_x>0$) vortex harmonics have the generation potential and
the linearly generated acoustic
wave harmonics are only down-shear tilted ($k_y/k_x<0$). Such a
topology can not be reproduced by the linear source term
representations of any classical formulation of an acoustic analogy
\citep{Hau2015}. The topological incompatibility of the linear
source terms of the latter and the linear
mechanism of wave generation induced by the flow non-normality is fundamental and, consequently,
raises righteous questions about the validity of interpretations arising from applying of
acoustic analogies to a shear flow systems.

\section{Acknowledgment}
This work was supported by the German Scientific Society (DFG)
research grant No.FO 674/6-1. We gratefully acknowledge support with computing time
from HPC facility at the university of Siegen.

\bibliography{Literature}

\end{document}